\documentclass[trackchanges,twocolumn]{aastex701}
\usepackage[T1]{fontenc}
\usepackage{ae,aecompl}

\usepackage{upgreek}
\usepackage{amsmath}	
\usepackage{amssymb}	

\usepackage{CJK}

\begin{document}
\begin{CJK*}{UTF8}{gkai}

\title{Accurately Measuring Variability Properties from Noisy Light Curves}

\email{zbsu@mail.ustc.edu.cn; zcai@ustc.edu.cn}

\author[orcid=0000-0001-8515-7338]{Zhen-Bo Su（苏镇波）}
\affiliation{Department of Astronomy, University of Science and Technology of China, Hefei 230026, People's Republic of China}
\affiliation{School of Astronomy and Space Science, University of Science and Technology of China, Hefei 230026, People's Republic of China}
\email[show]{zbsu@mail.ustc.edu.cn}  

\author[0000-0002-4223-2198]{Zhen-Yi Cai}
\affiliation{Department of Astronomy, University of Science and Technology of China, Hefei 230026, People's Republic of China}
\affiliation{School of Astronomy and Space Science, University of Science and Technology of China, Hefei 230026, People's Republic of China}
\email[show]{zcai@ustc.edu.cn}

\author[orcid=0009-0005-3916-1455,sname='Xia']{Ruisong Xia}
\affiliation{Department of Astronomy, University of Science and Technology of China, Hefei 230026, People's Republic of China}
\affiliation{School of Astronomy and Space Science, University of Science and Technology of China, Hefei 230026, People's Republic of China}
\email[hide]{xiars@mail.ustc.edn.cn}

\author[orcid=0009-0004-9950-9807,sname='Chen']{Shi-Jiang Chen}
\affiliation{Department of Astronomy, University of Science and Technology of China, Hefei 230026, People's Republic of China}
\affiliation{School of Astronomy and Space Science, University of Science and Technology of China, Hefei 230026, People's Republic of China}
\email[hide]{csj666@mail.ustc.edn.cn}

\author[0000-0003-4200-4432]{Lulu Fan (范璐璐)}
\affiliation{Department of Astronomy, University of Science and Technology of China, Hefei 230026, People's Republic of China}
\affiliation{School of Astronomy and Space Science, University of Science and Technology of China, Hefei 230026, People's Republic of China}
\affiliation{College of Physics, Guizhou University, Guiyang 550025, People's Republic of China}
\affiliation{Deep Space Exploration Laboratory, Hefei 230088, People's Republic of China}
\email[hide]{llfan@ustc.edu.cn}

\author[0000-0001-8416-7059]{Hengxiao Guo}
\affiliation{Shanghai Astronomical Observatory, Chinese Academy of Sciences, Shanghai 200030, People's Republic of China}
\email[hide]{hengxiaoguo@gmail.com}

\author[0000-0002-0771-2153]{Mouyuan Sun}
\affiliation{Department of Astronomy, Xiamen University, Xiamen 361005, People's Republic of China}
\email[hide]{msun88@xmu.edu.cn}

\author[0000-0002-4419-6434]{Jun-Xian Wang}
\affiliation{Department of Astronomy, University of Science and Technology of China, Hefei 230026, People's Republic of China}
\affiliation{School of Astronomy and Space Science, University of Science and Technology of China, Hefei 230026, People's Republic of China}
\affiliation{College of Physics, Guizhou University, Guiyang 550025, People's Republic of China}
\email[hide]{jxw@ustc.edu.cn}



\begin{abstract}
    Modern time-domain surveys probe optical variability near photometric error, challenging weak-signal recovery. The structure function (SF) is widely used for characterizing variability of active galactic nuclei (AGNs), but its statistical recovery limits are unclear. Using simulated light curves with realistic errors, we compare three SF estimators from the literature: two non-parametric (pairwise noise subtraction and Bayesian) and one parametric. All recover intrinsic SF when variability dominates, but diverge in the weak regime. 
    The pairwise noise-subtraction estimator approaches an artificial floor when negative variance estimates are discarded, with the floor set by photometric precision and sampling rather than intrinsic variability.
    The parametric method suffers degeneracy, 
    while the Bayesian estimator yields nearly unbiased SF estimates given well-calibrated photometric errors and adequate sampling.
    We also derive an idealized detection threshold linking intrinsic variability, photometric precision, and pair counts, providing a reference for assessing the detectability of weak signals.
    Our results establish statistical limits for SF measurements in the weak-variability regime and offer a framework for accurately measuring low-amplitude variability in AGNs (e.g., extremely stable quasars) and in new populations such as little red dots discovered by JWST.
\end{abstract}


\section{Introduction}
Active galactic nuclei (AGNs) are among the most luminous persistent sources in the universe, powered by accretion onto central black holes spanning $\sim10^4$--$10^9\,M_\odot$, from intermediate-mass black holes in dwarf galaxies to supermassive black holes in luminous quasars \citep{Padovani2017}. One of the defining properties of AGNs is their stochastic variability across the electromagnetic spectrum, on timescales ranging from hours to decades \citep[e.g.,][]{Ulrich1997}. Because the central engine cannot be spatially resolved for most AGNs, variability provides one of the few indirect probes of accretion physics \citep[e.g.,][]{Cai2018,Su2025}.

In the community, the variability is commonly quantified in two broad ways: by estimating the variability directly from the data, and by modelling the light curve with a stochastic process. The two most widely used direct measures, both adopted from time-series analysis, are the structure function (SF) and the power spectral density (PSD).
For UV/optical variability, the SF has traditionally been the primary tool for quantifying variability, as it is computed directly from magnitude differences and requires no assumption about the underlying process \citep[e.g.,][]{Hughes1992ApJ...396..469H,diClemente1996ApJ...463..466D,Kozlowski2016ApJ...826..118K}. 
The PSD, long the standard tool for X-ray variability studies where evenly sampled light curves are routinely available \citep[e.g.,][]{Uttley2002,Vaughan2003MNRAS}, has more recently been applied in the optical as well, either from the continuous sampling of space telescopes \citep[e.g.,][]{Mushotzky2011ApJ...743L..12M,Kasliwal2015,Yuk2025} or, for irregularly sampled ground-based data, using methods developed for uneven sampling \citep[e.g.,][]{Lomb1976,Scargle1982,arevalo2012,Petrecca2025}. 
The alternative is to model the light curve with a stochastic process, such as the damped random walk \citep[DRW;][]{Kelly2009ApJ...698..895K} or the more general continuous-time autoregressive moving-average (CARMA) process \citep{Kelly2014}. Gaussian-process \citep[GP;][]{celerite} methods provide a flexible and efficient framework for fitting such stochastic models with a variety of covariance kernels \citep{Lefkir2025,Yu2022,Yu2025}.

Together, these approaches have established a broadly consistent picture of optical variability, though almost exclusively for luminous quasars. Variability amplitudes increase toward longer timescales, consistent with a stochastic red-noise process \citep[e.g.,][]{Giveon1999,Collier2001}, while stochastic-process models such as the DRW provide an adequate description on month-to-year timescales \citep[e.g.,][]{Kelly2009ApJ...698..895K,MacLeod2010ApJ...721.1014M,MacLeod2012ApJ...753..106M,Zu2013,2025MNRAS.544L..96X}.

This consensus, however, was established over a limited range of timescales. 
As time-domain surveys have expanded both their temporal baselines and cadence coverage, AGN variability can now be studied over a wider range of timescales. Multi-survey studies have combined data from several optical surveys, including the Sloan Digital Sky Survey (SDSS, \citealt{York2000}), Pan-STARRS1 (PS1, \citealt{Chambers2016}), and the Zwicky Transient Facility (ZTF, \citealt{Bellm2019}), to extend AGN variability measurements to multi-decade baselines (e.g., \citealt{Stone2023,Arevalo2024,Yu2025}), while high-cadence observations probe shorter timescales (e.g., \citealt{Tang2024,Yuk2025}). Variability measurements now extend beyond the traditional month-to-year regime, and departures from the simple DRW model have begun to emerge.

For example, the DRW predicts an SF that approaches a plateau at timescales longer than its characteristic timescale, $\tau_{\rm d}$, corresponding to a flattening of the PSD at low frequencies. However, multi-decade monitoring has revealed departures from this expectation, likely associated with excess low-frequency power beyond the DRW model \citep[e.g.,][]{Guo2017ApJ...847..132G,Petrecca2025,Arevalo2024,Yuk2025}.
At the opposite extreme, high-cadence observations from \textit{Kepler} and other surveys probe variability on timescales $\Delta t \ll \tau_{\rm d}$, where the DRW predicts ${\rm SF}\propto\Delta t^{1/2}$ and a PSD proportional to $f^{-2}$, with $\Delta t$ denoting the variability timescale and $f$ the temporal frequency.
Although some studies report deviations from the DRW prediction, including steeper short-timescale SFs or high-frequency PSDs  \citep[e.g.,][]{Mushotzky2011ApJ...743L..12M,Caplar2017,Stone2023,Yuk2025}, others find that the DRW remains consistent with the data \citep[e.g.,][]{Tang2024,Yuk2026}. 
The origin of this discrepancy remains unclear. Unlike the long-timescale regime, short-timescale measurements increasingly approach the photometric-noise limit. For a red-noise process, the intrinsic variability signal decreases toward shorter timescales. As a result, the observed fluctuations have lower signal-to-noise ratios, and the recovery of intrinsic variability becomes increasingly sensitive to the characterization and subtraction of photometric errors.

The same difficulty becomes more pronounced when moving from luminous quasars to fainter AGNs.
Only recently has attention turned to lower-mass and lower-luminosity systems \citep[e.g.,][]{Burke2020ApJ...899..136B,Zuo2024,Su2024a,Beard2025,Tan2026,Zuo2026}, where weaker variability signals and larger relative photometric errors make intrinsic variability increasingly difficult to recover.
A striking example is the nucleus of the dwarf galaxy NGC 4395, for which the same light curve \citep{McHardy2023MNRAS.519.3366M} has led to conflicting interpretations of variability on timescales $\lesssim$ hours. 
Using PSRESP modelling of the PSD \citep{Uttley2002}, \citet{Beard2025} found that both the DRW and models with steeper PSD slopes provide acceptable descriptions, whereas \citet{Su2024a} found the measured SF to be inconsistent with the DRW prediction. 
This discrepancy may reflect the difficulty of recovering weak variability signals when the intrinsic fluctuations approach the photometric-noise level, as well as the non-trivial connection between PSD- and SF-based variability characterizations under realistic observing conditions, where sampling, noise, and estimator effects can differentially influence the recovered variability properties \citep[e.g.,][]{Sartori2019}.
Regardless of its origin, this case illustrates that in the weak-variability regime, conclusions about the underlying stochastic process may depend critically on how intrinsic variability is extracted from noisy observations.

The challenge becomes even greater for sources whose intrinsic variability is so weak that establishing whether they vary at all becomes the primary question. For such sources, the detection of variability depends critically on how reliably intrinsic signals can be separated from photometric error.
The little red dots \citep[LRDs; e.g., ][]{Labbe2023,Matthee2024,Kocevski2023} recently discovered by James Webb Space Telescope (JWST) represent one such case. Their reported variability amplitudes are often comparable to or below the photometric errors. For a large fraction of these objects, only upper limits rather than significant detections are available, leaving whether they exhibit AGN-like variability unresolved \citep[e.g.,][]{Zhang+2025,Stone2025,Zhou2025,Secunda2026}.

Similarly, extremely stable quasars \citep[ESQs;][]{Kang2024ApJ...971...60K} are type 1 quasars showing extremely weak optical variability over a decade of monitoring. Although the long-term variation amplitudes of ESQs are clearly smaller than those of ordinary quasars, whether their variability amplitudes increase with timescale in the same way remains unclear and can only be resolved with reliable measurements of variability near the photometric error level.

Together, these examples represent progressively more demanding tests of our ability to recover intrinsic variability from noisy data. Addressing this challenge requires statistical methods that can separate intrinsic fluctuations from photometric errors.
In the frequency domain, various approaches have been developed to recover intrinsic PSDs despite observational effects, such as the improved periodogram estimates \citep{Papadakis1993}, forward-modelling or leakage-based corrections \citep[e.g.,][]{Uttley2002,ZhuSF2016} and Gaussian-process modelling \citep[e.g.,][]{Kelly2014,celerite,Lefkir2025,Yu2026}. 
However, model-based approaches require assumptions about the underlying spectral or covariance structure, and a good fit does not by itself establish that the assumed process is the true one \citep{Kozlowski2016b}.

The SF provides a complementary time-domain approach by directly characterizing the covariance structure without assuming a specific variability model, as advocated by \citet{Kozlowski2016b}. 
Compared with Fourier-based methods, it is less directly affected by spectral leakage and aliasing, and its non-parametric nature makes it attractive for probing stochastic variability.  However, model independence does not guarantee reliable recovery of the intrinsic variability signal. \citet{Emmanoulopoulos2010MNRAS.404..931E} showed that spurious breaks can appear in SFs even when the underlying process contains no characteristic timescale, while \citet{Kozlowski2016ApJ...826..118K} found that commonly used SF estimators may incompletely account for photometric errors and produce artificially flattened slopes. On short timescales, the observed SF can become dominated by measurement noise, making reliable recovery of the intrinsic signal dependent on accurate noise treatment.

Several approaches have been proposed to remove noise from SF measurements. Two are non-parametric and make no assumptions about the SF shape: the pair-wise noise subtraction method of \citet{Press1992} removes the known photometric errors from individual pairs, while the Bayesian estimator of \citet{Suberlak2017} infers the intrinsic scatter within each timescale bin from the distribution of magnitude differences. A third, the parametric method of \citet{Kozlowski2016ApJ...826..118K}, jointly models the variability signal and photometric error. These approaches rely on fundamentally different assumptions, yet their ability to recover intrinsic SFs—and the statistical limits of that recovery—in the weak-signal regime have not been systematically established.

In this paper we quantify how reliably the intrinsic SF can be recovered from noisy light curves, and what sets the limit of that recovery. Section~\ref{sec:sim_lc} describes the simulated light curves and the adopted photometric error model. Section~\ref{sec:measure_sf} introduces the three SF estimators and compares their performance as the intrinsic amplitude falls below the photometric error. Section~\ref{sec:discussion} examines why each estimator fails, extends simulations to considering realistic sampling, to non-DRW processes, and to biased photometric errors, and finally applies the three estimators to NGC~4395. Section~\ref{sec:conclusion} summarises our conclusions.

\section{AGN Light Curves} \label{sec:sim_lc}

\subsection{Ideal light curves}

\begin{figure*}
    \centering
    \includegraphics[width=0.98\linewidth]{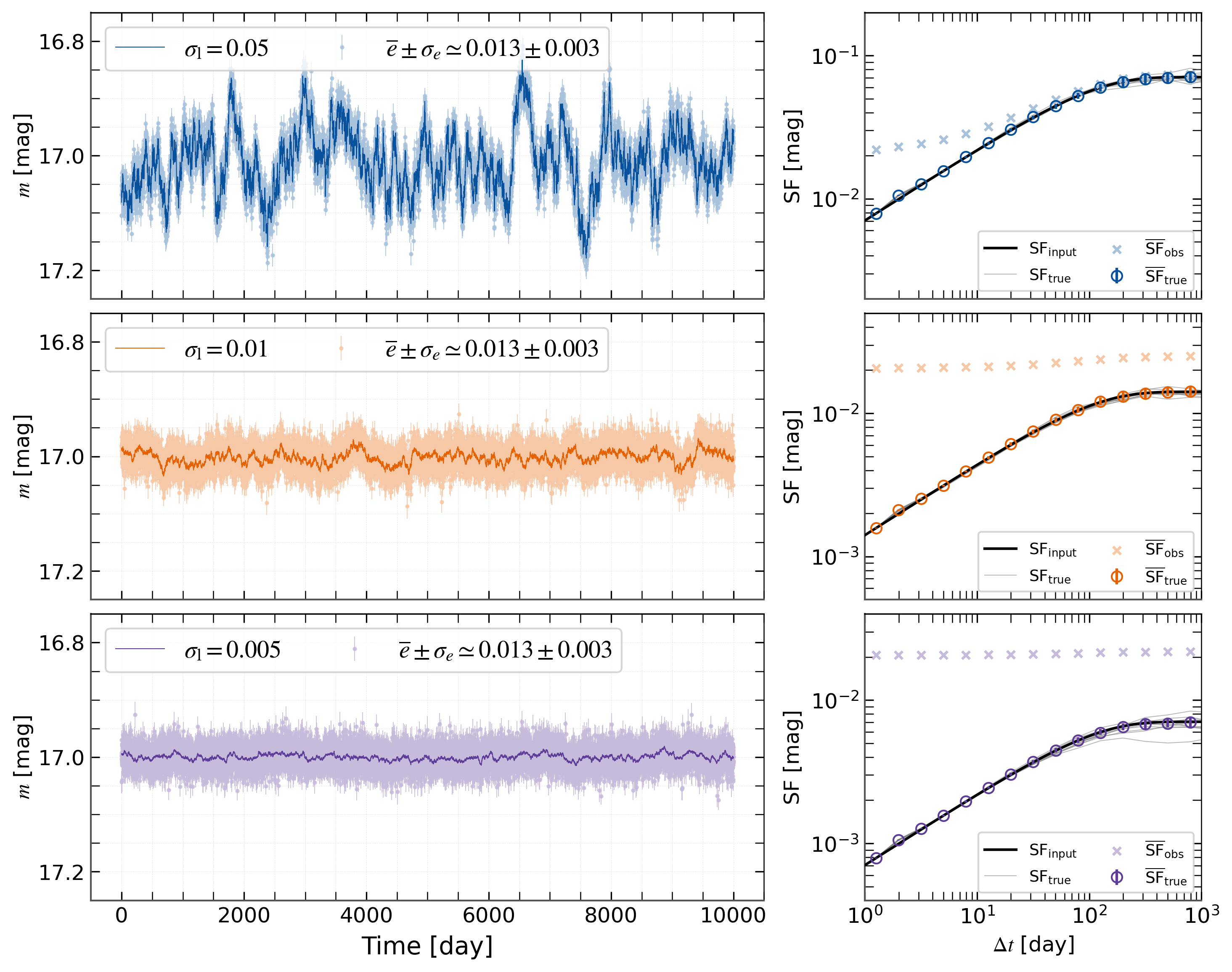}
    \caption{Illustration of simulated light curves (Section~\ref{sec:sim_lc}) and SFs (Section~\ref{sec:measure_sf}) for AGNs with mean magnitude $\bar m = 17$~mag. 
    \textit{Left panels:} From top to bottom, the three thick curves show randomly selected ideal light curves with $\sigma_{\rm l} = 0.05$, 0.01, and 0.005~mag, respectively. The corresponding mock AGN light curves are plotted as small points with error bars, after including photometric errors with a mean of $\bar e \simeq 0.013$~mag and a standard deviation of $\sigma_e = 0.003$~mag (Figure~\ref{fig:e_sigma_e_mag}). 
    \textit{Right panels:} Thick black curves are the input SFs used to generate the simulated light curves. Thin gray curves show the true SFs of 10 ideal realizations. Circles (crosses) mark the median SF of 100 true (observed) SFs, with error bars indicating the 16$-$84th percentile ranges. 
    }
    \label{fig:ideal-lc-sf}
    
\end{figure*}

Following \citeauthor{Press1992} (\citeyear{Press1992}; see also \citealt{Zu2011,Kozlowski2016b}), we simulate an ideal AGN light curve in magnitude as a realization of a stationary stochastic process, whose temporal correlation is described by a power-exponential covariance function,
\begin{align}
{\rm cov}(\Delta t \mid \sigma_{\rm l},\tau_{\rm d},\beta) &\equiv \sigma_{\rm l}^2 {\rm ACF}(\Delta t | \tau_{\rm d}, \beta) \nonumber \\
&= \sigma_{\rm l}^2
\exp\left[-\left(\frac{\Delta t}{\tau_{\rm d}}\right)^{\beta}\right],
\label{eq:cov}
\end{align}
where $\Delta t = |t_j - t_k|$ is the time interval between two epochs at times $t_j$ and $t_k$, $\tau_{\rm d}$ is the characteristic decorrelation timescale, $\sigma_{\rm l}^2$ is the long-term ($\Delta t \gg \tau_{\rm d}$) variance, ${\rm ACF}(\Delta t | \tau_{\rm d}, \beta)$ is the normalized autocovariance function of light curve, and $\beta$ determines the timescale dependence of the short-term ($\Delta t \ll \tau_{\rm d}$) variation amplitude.

As a reference, we adopt $\tau_{\rm d} = 100$~d, $\sigma_{\rm l} = 0.05$ mag, and $\beta = 1$, corresponding to a DRW process. These selections are guided by observations of typical AGNs with massive black holes \citep[e.g.,][]{MacLeod2010ApJ...721.1014M,Suberlak2021ApJ...907...96S,Burke2021Sci...373..789B}. 
We also consider two smaller values for $\sigma_{\rm l}$, i.e., $\sigma_{\rm l} = 0.01$~mag and 0.005~mag, for reasons addressed in Section~\ref{subsec:parameter_selection}.

For a given set of variation parameters, $\{\tau_{\rm d},~\sigma_{\rm l},~\beta\}$, we simulate an ideal AGN light curve, $\{t_j,m_j\}_{j=1}^{N_{\rm ep}}$, where $N_{\rm ep}$ is the number of observing epochs and $m_j$ is the ideal magnitude at epoch $t_j$.
We adopt a uniform fine cadence of $\mathcal{C} = 0.5$~d and a maximum baseline of $\mathcal{T} = 10^4$~d, spanning a very broad dynamical range of timescales from $0.005\tau_{\rm d}$ to $100\tau_{\rm d}$ for fiducial $\tau_{\rm d} = 100$~d.
Given $\mathcal{C}$ and $\mathcal{T}$, we have a series of $\{\Delta t_i\}$ from $\Delta t_1 = \mathcal{C}$ to $\Delta t_{N_{\rm ep}} = \mathcal{T}$ in steps of $\mathcal{C}$, and then construct a $N_{\rm ep} \times N_{\rm ep}$ covariance matrix $\boldsymbol{C}$ using Equation~(\ref{eq:cov}). Following \citet{Kozlowski2016b}, we decompose $\boldsymbol{C}$ as $\boldsymbol{C} = \boldsymbol{L}^{\rm T}\boldsymbol{L}$, where $\boldsymbol{L}$ is an $N_{\rm ep} \times N_{\rm ep}$ matrix.\footnote{We use singular value decomposition (SVD) rather than the Cholesky decomposition used by \citet{Kozlowski2016b}, since dense sampling can make the covariance matrix numerically ill-conditioned, particularly for $\beta=2$, causing Cholesky decomposition to fail.}
Multiplying $\boldsymbol{L}^{\rm T} $ by a $N_{\rm ep} \times 1$ matrix $\boldsymbol{r}$ whose elements are randomly drawn from a standard normal distribution, i.e., $\boldsymbol{y} = \boldsymbol{L}^{\rm T} \boldsymbol{r}$, gives rise to an ideal AGN light curve, $\boldsymbol{y}$, with zero mean magnitude. 
By repeating random realizations of $\boldsymbol{r}$, we generate 100 distinct ideal AGN light curves. Following \citet{Kozlowski2016b}, we assign each light curve a mean magnitude of $\bar m=17$~mag as our fiducial setting.
The dark thick curves in the left panels of Figure~\ref{fig:ideal-lc-sf} illustrate representative ideal AGN light curves for $\sigma_{\rm l}=0.05$, 0.01, and 0.005 mag from top to bottom, respectively.

\subsection{Mock light curves}\label{subsec:real_lc}

The unavoidable photometric error prevents us from digging into the intrinsic AGN variability especially at short timescales. To mimic the effect of photometric error, we add random Gaussian deviates to all magnitudes in an ideal AGN light curve and generate a corresponding mock AGN light curve, $\{t_j,m_j',e_j\}$, where $m'_j$ is the observed magnitude given a photometric error of $e_j$ at epoch $t_j$. Here, $m'_j = m_j + \varepsilon_j$ and $\varepsilon_j \sim \mathcal{N}(0, e_j^2)$, where ${\mathcal{N}}(0, e_j^2)$ is a Gaussian distribution with a mean of zero and a variance of $e_j^2$. 

\begin{figure}
    \centering
    \includegraphics[width=0.95\linewidth]{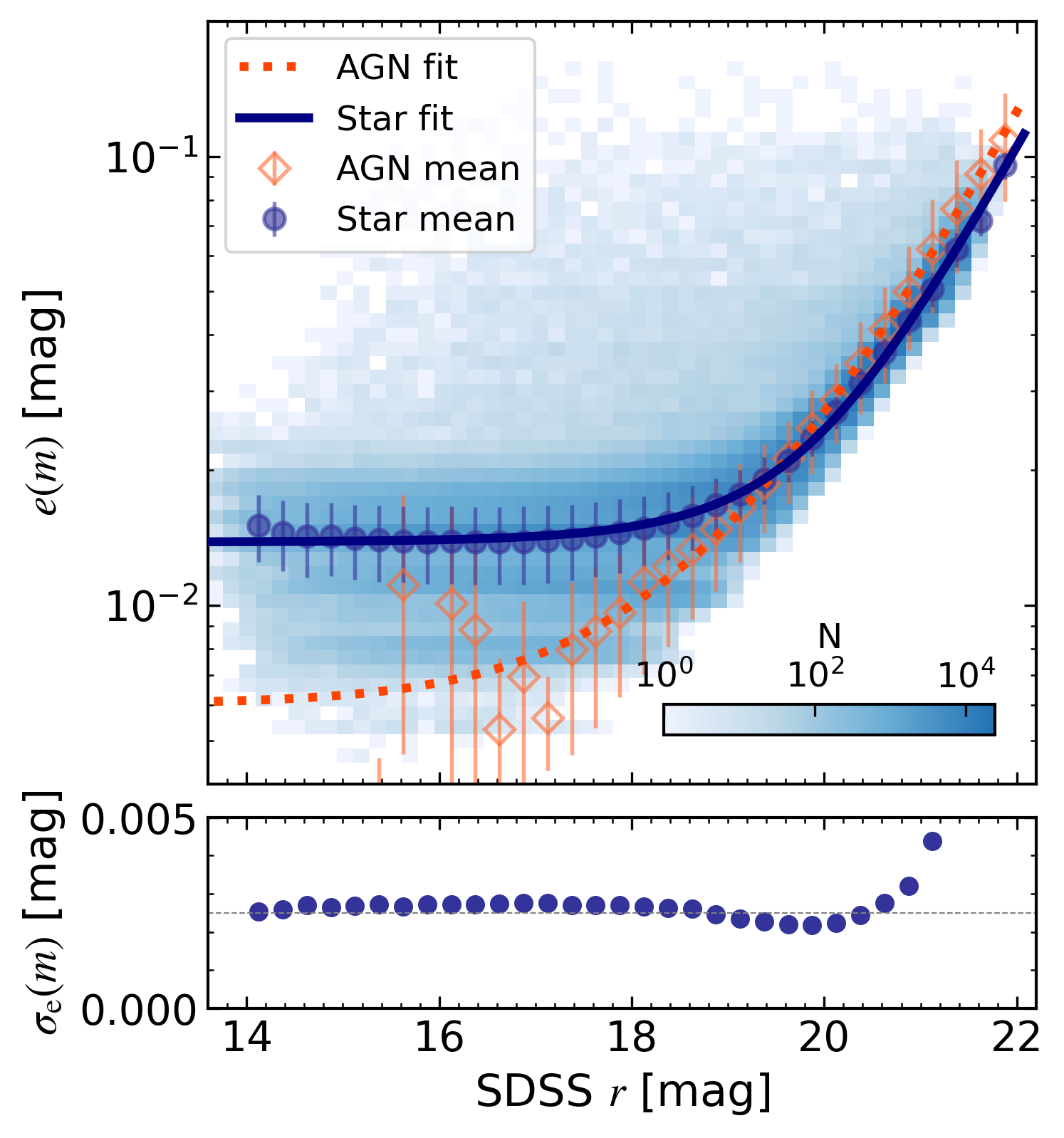}
    \caption{Photometric error properties of a modern SDSS-like telescope. 
    Top panel: photometric errors as a function of $r$-band magnitude for non-variable standard stars \citep[colored background shows the distribution of stellar photometric errors;][]{Ivezic2007} and for AGNs in the Stripe 82 \citep[][]{MacLeod2010ApJ...721.1014M}. 
    Circles (diamonds) with error bars indicate the mean and standard deviation of photometric errors of stars (AGNs) in magnitude bins, with an overlaid solid (dotted) curve representing an analytic error model.
    Bottom panel: standard deviation of stellar photometric errors as a function of magnitude.}
    \label{fig:e_sigma_e_mag}
\end{figure}

In reality, $e_j(m)$ depends on the magnitude and varies even for the same magnitude. The color background in the main panel of Figure~\ref{fig:e_sigma_e_mag} illustrates the distribution of photometric error versus magnitude for a sample of well-calibrated non-variable standard stars in Stripe 82\footnote{Stripe 82 is an equatorial region repeatedly imaged by the Sloan Digital Sky Survey \citep[SDSS;][]{Abazajian2009}.} (\citealt{Ivezic2007}; catalog v2.6~\footnote{\url{https://faculty.washington.edu/ivezic/sdss/catalogs/stripe82.html}}). 
We note that an updated catalog, incorporating additional post-2007 SDSS imaging and recalibration based on Gaia EDR3, is now available \citep{Thanjavur2021}. However, we use here version 2.6 of the catalog to allow a consistent comparison with the Stripe 82 AGN photometry of \citet{MacLeod2010ApJ...721.1014M}, as shown in Figure~\ref{fig:e_sigma_e_mag}. 
Since the catalog only provides mean magnitudes rather than individual light curves for the standard stars, we estimate a characteristic single-epoch photometric error, $e_\star$, from the reported standard error of the mean magnitude, $\sigma_{\rm mean}$. This estimate assumes that the intrinsic variability of these standard stars is negligible compared with the observational noise and that the single-epoch photometric uncertainty for a given star remains approximately constant among epochs. 
Assuming further that the measurement errors are independent between epochs, the standard error of the mean is approximately $\sigma_{\rm mean}\simeq e_\star/\sqrt{N_{\rm obs}}$, where $N_{\rm obs}$ is the number of epochs. We therefore estimate $e_\star\simeq\sigma_{\rm mean}\sqrt{N_{\rm obs}}$ where both $\sigma_{\rm mean}$ and $N_{\rm obs}$ are reported in the catalog.
We take the SDSS $r$ band as a reference and, from $r = 14$~mag to 22~mag in bins of 0.25~mag, obtain the mean photometry errors ($\bar e = \langle e_\star \rangle$) and the associated standard deviations ($\sigma_e$), shown as circles with error bars in the main panel of Figure~\ref{fig:e_sigma_e_mag}.

Following \citet{Ivezic2019} and \citet{Suberlak2021ApJ...907...96S}, we use an analytic error model to characterize the dependence of $\bar e$ on magnitude, that is, ${\bar e}^2(m) = \sigma_{\rm sys}^2 + \sigma_{\rm rand}^2(m)$, where the systematic term ($\sigma_{\rm sys}$) is independent of magnitude and the statistical term ($\sigma_{\rm rand}$) depends on magnitude as $\sigma_{\rm rand}^2 = (0.04 - \gamma)\,x + \gamma x^2$ with $x = 10^{0.4(m-m_5)}$. Here, $m_5$ is the 5$\sigma$ limiting magnitude for a point source in a given band, and $\gamma$ is a parameter that depends on the sky brightness and other observing conditions \citep{Ivezic2019}. 
By fitting this error model to the binned photometric errors of standard stars, we find $\sigma_{\rm sys} = 0.0138 \pm 0.0001$, $\gamma = 0.0379 \pm 0.0003$, and $m_5 = 22.74 \pm 0.02$ for the SDSS $r$-band photometry (solid curve in the main panel of Figure~\ref{fig:e_sigma_e_mag}).

Using the above error model, we assign a mock observed photometric error $e_j$ to a given magnitude $m_j$ as $e_j(m_j) = \bar{e}(m_j) + \delta_j$ with $\delta_j \sim \mathcal{N}(0, \sigma_e^2)$. We take $\sigma_e = 0.003$~mag guided by the fact that $\sigma_e$ depends weakly on magnitude as illustrated in the bottom panel of Figure~\ref{fig:e_sigma_e_mag}. Note that for $m_j \simeq 17$~mag the corresponding $\bar e \simeq 0.013$~mag.
In the left panels of Figure~\ref{fig:ideal-lc-sf}, light small points with error bars show the mock light curves for AGNs with $r$-band $\bar m = 17$~mag. 
Since the median $r$-band magnitude of quasars in the SDSS Stripe 82 sample is $\simeq20$ mag \citep{MacLeod2010ApJ...721.1014M}; we therefore also consider a fainter-source case with $\bar m=20$ mag. 
The adopted error model gives $\bar e\simeq0.025$ mag at this magnitude, nearly twice the uncertainty in our fiducial $\bar m=17$ mag case. 
For fixed intrinsic variability, this lowers the signal-to-noise ratio $\sigma_{\rm l}/\bar e$ by roughly a factor of two. 
We examine how this change affects SF recovery in Section~\ref{subsec:photometric_error}; unless stated otherwise, the simulations use $\bar m=17$ mag.

\subsection{Simulation parameters}\label{subsec:parameter_selection}

To simulate AGN light curves, we need to set up many simulation parameters, i.e., $\{\tau_{\rm d},~\sigma_{\rm l},~\beta,~\mathcal{C},~\mathcal{T},~\bar m,~\bar e,~\sigma_e\}$.
We note that the aforementioned specific choices of these simulation parameters do not weaken our demonstration of determining the best SF estimator (Section~\ref{sec:measure_sf}).

Although both $\sigma_{\rm l}$ and $\bar e$ may depend on magnitude, the ratio of $\sigma_{\rm l}$ to $\bar e$, rather than the absolute value of either $\sigma_{\rm l}$ or $\bar e$, is critical in determining the best SF estimator. 
If $\sigma_{\rm l} > \bar e$, e.g., $\sigma_{\rm l} / \bar e \simeq 4$ (top-left panel of Figure~\ref{fig:ideal-lc-sf}), the AGN variability is significant and the effect of photometric error is minor except at the very short timescales.
Once $\sigma_{\rm l} \simeq \bar e$, e.g., $\sigma_{\rm l} / \bar e \simeq 0.8$ (middle-left panel of Figure~\ref{fig:ideal-lc-sf}), the AGN variability is marginal and the photometric error is troublesome at nearly all timescales.
Even worse, the AGN variability is completely overwhelmed by the photometric error when $\sigma_{\rm l} < \bar e$, e.g., $\sigma_{\rm l} / \bar e \simeq  0.4$ (bottom-left panel of Figure~\ref{fig:ideal-lc-sf}). 
Therefore, fixing $\bar e \simeq 0.013$, considering different values of $\sigma_{\rm l}$ is essential for determining the best SF estimator.

\begin{figure}
    \centering
    \includegraphics[width=0.98\linewidth]{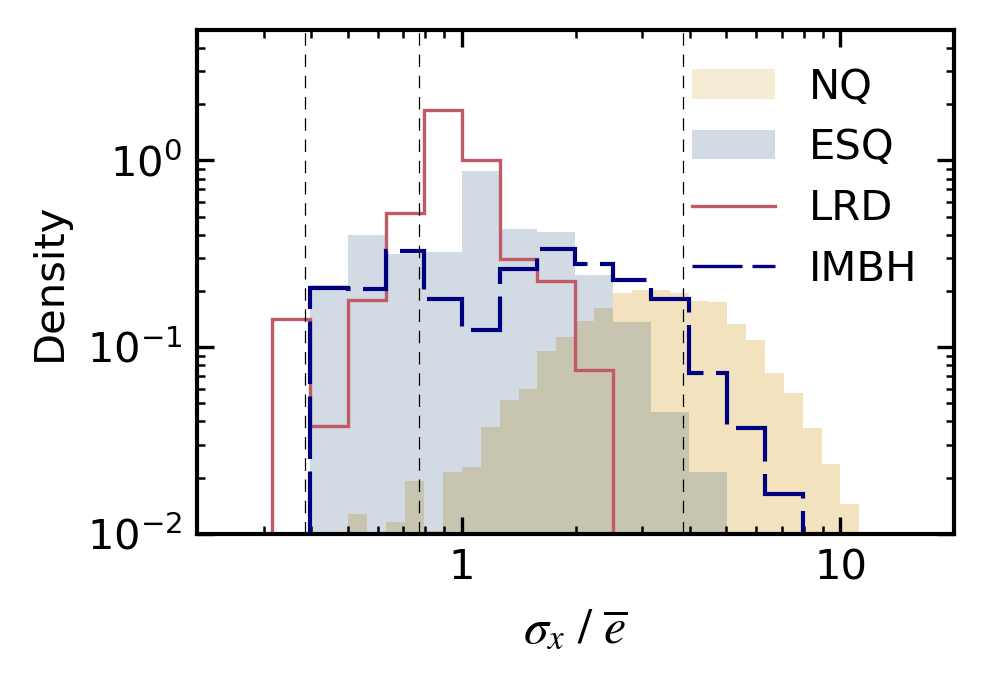}
    \caption{Distributions of the ratio of variation amplitude to mean photometric error, $\sigma_x / \bar e$, for normal quasars \citep[NQs:][]{MacLeod2010ApJ...721.1014M}, extremely stable quasars \citep[ESQs;][]{Kang2024ApJ...971...60K}, little red dots \citep[LRDs;][]{Zhang+2025}, and intermediate-mass black holes (IMBHs; \citealt{Guo2026arXiv260902179G}). 
    We include SDSS $r$-band light curves for NQs and ESQs, all available JWST light curves for LRDs, and ZTF $r$-band light curves for IMBHs. 
    The baseline is $\simeq 10$~yr in the observed frame for NQs and ESQs, $\simeq 5$~yr for IMBHs, but only $\lesssim 2$~yr for LRDs. The three vertical dashed lines (right to left) correspond to $\sigma_{\rm l} = 0.05$, 0.01, and 0.005~mag, assuming $\bar e = 0.013$~mag.}
    \label{fig:dis_sigma_pe}
\end{figure}

To demonstrate that considering the three values of $\sigma_{\rm l}$ relative to $\bar e$ is sufficient, Figure~\ref{fig:dis_sigma_pe} illustrates distributions of variation amplitude to mean photometric error, $\sigma_x/\bar e$, for four samples: normal quasars \citep[NQs:][]{MacLeod2010ApJ...721.1014M}, extremely stable quasars \citep[ESQ;][]{Kang2024ApJ...971...60K},  little red dots \citep[LRDs;][]{Zhang+2025}, and intermediate-mass black holes (IMBHs; \citealt{Guo2026arXiv260902179G}).
For a light curve of a source, we estimate the variation amplitude, $\sigma_x$, and the mean photometric error, $\bar e$.
Here, $\sigma_x$ quantifies the intrinsic scatter in magnitude after subtracting the contribution from photometric noise, using the square root of the weighted excess variance introduced by \citet{Kang2024ApJ...971...60K}. The subscript $x$ indicates that the amplitude can be estimated in different bands or over distinct baselines. For example, Figure~\ref{fig:dis_sigma_pe} uses the full light curve to estimate $\sigma_x$, whereas Figures~\ref{fig:p92_vs_k16} and \ref{fig:p92_vs_s17} use the timescale-dependent variation amplitude $\sigma_x(\Delta t) = {\rm SF}_{\rm input}(\Delta t, \sigma_{\rm l}, \tau_{\rm d}, \beta) /\sqrt{2}$ for each timescale bin (see Section~\ref{sec:measure_sf}).

All NQs and ESQs are located in the SDSS Stripe~82. \citet{Kang2024ApJ...971...60K} showed that a unique population of 136 ESQs with extremely weak variability in UV/optical over $\simeq 10$~yr has a prominent excess of radio emission, thus posing a new clue to jet launching. In Figure~\ref{fig:dis_sigma_pe} we show the distribution of $\sigma_x/\bar e$ for 136 ESQs using their $r$-band light curves with at least 10 physical photometric measurements. Variabilities of $\simeq 6\%$ ESQs are statistically non-detected and their $\sigma_x$ are replaced by $2\sigma$ upper limits following \citet{Kang2024ApJ...971...60K}.
For comparison, we also show the distribution of $\sigma_x / \bar e$ for 9004 NQs selected from the 9258 spectroscopically confirmed quasars \citep{MacLeod2010ApJ...721.1014M} by requiring at least 10 physical photometric measurements in the $r$-band light curve and excluding ESQs.

\cite{Zhang+2025} reported no strong variability for a sample of 314 LRD candidates with multi-epoch JWST observations, including NIRcam and MIRI imaging data. 
In total, there are 1286 LRD light curves over $\lesssim 2$~yr possessing two to ten photometric measurements with a median of three\footnote{\url{https://github.com/Zijian-astro/variable_LRD_candidates/}}. 
We focus on 258 LRD light curves with at least five photometric measurements.
Figure~\ref{fig:dis_sigma_pe} also shows the resulting distribution of $\sigma_x / \bar e$ for LRDs, among which $\simeq 60\%$ are $2\sigma$ upper limits.

\citet{Guo2026arXiv260902179G} compiled 192 reliable IMBH candidates at $z\lesssim0.3$ with $M_{\rm BH}<10^6\,M_\odot$. For these candidates, we retrieved public $r$-band light curves spanning $\simeq 5$~yr from the Zwicky Transient Facility (ZTF; \citealt{Bellm2019, Masci2019}), using PSF-fitting photometry from the \texttt{Zubercal} service\footnote{\url{http://nesssi.cacr.caltech.edu/ZTF/Web/Zuber.html}}, with a matching radius of $2''$. 
We retained only epochs with ${\tt catflags}=0$ to exclude measurements affected by known image-quality or photometric-quality issues. This selection resulted in 189 sources with at least 10 good measurements. Their distribution of $\sigma_x/\bar e$ is also shown in Figure~\ref{fig:dis_sigma_pe}.

\section{Structure Functions}\label{sec:measure_sf}

The input covariance function, Equation~(\ref{eq:cov}), corresponds to an input SF of
\begin{align}\label{eq:sf_input}
{\rm SF}_{\rm input}&(\Delta t \mid {\rm SF}_\infty,\tau_{\rm d},\beta) \nonumber\\
&= \sqrt{2 \sigma^2_{\rm l} - 2 {\rm cov}(\Delta t \mid \sigma_{\rm l},\tau_{\rm d},\beta)} \nonumber \\
&= {\rm SF}_\infty \sqrt{1 - \exp{\left[-\left(\frac{\Delta t}{\tau_{\rm d}}\right)^\beta\right]}},
\end{align}
where ${\rm SF}_\infty = \sqrt{2} \sigma_{\rm l}$ is another form of the long-term standard deviation. SF provides an intuitive, non-parametric characterization of variability as a function of timescale. 
The black thick solid curves in the right panels of Figure~\ref{fig:ideal-lc-sf} are the input SFs for our fiducial variation parameters. 
According to Equation~\eqref{eq:cov}, for a stationary process, Equation~\eqref{eq:sf_input} relates the SF to the ACF through $\mathrm{SF}_{\rm input}^2(\Delta t)=2\sigma_1^2[1-\mathrm{ACF}(\Delta t)]$. 
Given regularly sampled time series, residual autocorrelation tests, such as the Durbin--Watson \citep{DURBINTests}, Ljung--Box \citep{LjungTests}, and Breusch--Godfrey \citep{Breusch1978Testing,godfrey1978testing} tests, can help assess the adequacy of fitting statistical models (e.g., CARMA) to light curves.
However, these tests cannot be applied directly to SF estimates; here, we assess intrinsic SF recovery using simulated light curves.

In practice, given an ideal light curve, $\{t_j,m_j\}$ with $j=1,2,...,N_{\rm ep}$, the true SF at a timescale $\Delta t$ between $\Delta t_{\rm min}$ and $\Delta t_{\rm max}$ is generally measured according to
\begin{align}\label{eq:sf}
{\rm SF}_{\rm true}(\Delta t) &= \sqrt{\langle (\Delta m_i)^2 \rangle} \nonumber\\
&= \sqrt{\frac{1}{N_{\rm pair}} \sum_{i=1}^{N_{\rm pair}} (\Delta m_i)^2}  \quad {\rm mag},
\end{align}
where $\Delta t \equiv \sqrt{\Delta t_{\rm min} \Delta t_{\rm max}}$ is the center of a selected time bin from $\Delta t_{\rm min}$ to $\Delta t_{\rm max}$, and the average $\langle ... \rangle$ is taken over all $N_{\rm pair}$ pairs of $\Delta m_i = m_j - m_k$ with $\Delta t_i = t_j - t_k$ satisfying $\Delta t_{\rm min} \leq \Delta t_i <\Delta t_{\rm max}$.
Note that if the distribution of $\Delta m_i$ is Gaussian, Equation~(\ref{eq:sf}) is equivalent to $\sqrt{\frac{\pi}{2}}\langle |\Delta m_i| \rangle$ \citep[e.g.,][]{Hook1994MNRAS.268..305H,diClemente1996ApJ...463..466D} or $0.741 \times {\rm IQR}$, where ${\rm IQR}$ is the interquartile range between 25\% and 75\% of the distribution of $\Delta m_i$ \citep{MacLeod2012ApJ...753..106M}.

Throughout, we measure the SF over a dynamical range of timescales from 1 day to $10^3$ days (i.e., $0.01\tau_{\rm d}$ to $10\tau_{\rm d}$) in logarithmic steps of 0.2 dex.
As shown in each right panel of Figure~\ref{fig:ideal-lc-sf}, ${\rm SF}_{\rm true}$ of each ideal realization (a gray thin curve) is not exactly the same as ${\rm SF}_{\rm input}$ (the black thick curve) once $\Delta t \gtrsim \tau_{\rm d}$. This is simply owing to the randomness of AGN variability. The difference between ${\rm SF}_{\rm true}$ and ${\rm SF}_{\rm input}$ is visible even we have simulated light curves with an extremely long baseline of $100\tau_{\rm d}$. We expect smaller differences among ${\rm SF}_{\rm true}$ of ideal realizations, getting closer to ${\rm SF}_{\rm input}$, for longer baselines as indicated by ${\rm {\overline {SF}}}_{\rm true}$ for the median of 100 true SFs (open circles in the right panels of Figure~\ref{fig:ideal-lc-sf}).
Note that these finite-realization effects also warrant caution when interpreting SF features. \citet{Emmanoulopoulos2010MNRAS.404..931E} showed that even when the underlying process has no characteristic timescale there are SF breaks at long timescales mimicking artificial characteristic timescales as long as light curves are not long enough.
Instead, \citet{Sartori2019} showed that such long-timescale distortions are smoothed in a representative or ensemble SF, although finite light-curve length and discrete sampling can still affect the SF shape. 
\citet{Sartori2019} therefore restricted their analysis to timescales shorter than one-tenth of the total light-curve length (see also \citealt{Kozlowski2017A&A...597A.128K}). 
Our simulated light curves span $10^4$ days, and we examine timescales only up to $10^3$ days, following the same criterion. 
Indeed, the median ${\rm SF}_{\rm true}$ (circles) from 100 realizations is close to the input SF (black curves) over the timescales considered (right panels of Figure~\ref{fig:ideal-lc-sf}). 
At the short timescales of particular interest here, however, the intrinsic variability signal is weak and more strongly affected by photometric errors, making its reliable recovery particularly challenging, as we discuss below.

Similarly, we can get the observed SF from a mock AGN light curve, $\{t_j,m_j',e_j\}$, by
\begin{equation}\label{eq:sf_obs}
{\rm SF}_{\rm obs}(\Delta t) = \sqrt{\frac{1}{N_{\rm pair}} \sum_{i=1}^{N_{\rm pair}} (\Delta m_i')^2}  \quad {\rm mag},
\end{equation}
where $\Delta m_i' = m_j' - m_k'$. 
Also shown in each right panel of Figure~\ref{fig:ideal-lc-sf}, ${\rm SF}_{\rm obs}$ of mock realizations including photometric error (crosses for ${\rm {\overline {SF}}}_{\rm obs}$ being the median of 100 observed SFs) are distinguishable from ${\rm SF}_{\rm true}$ and ${\rm SF}_{\rm input}$. Given the same $\bar e$, departures of ${\rm SF}_{\rm obs}$ from ${\rm SF}_{\rm true}$ expand from short timescales in the case of $\sigma_{\rm l} > \bar e$ to nearly all timescales in the cases of $\sigma_{\rm l} \simeq \bar e$ and $\sigma_{\rm l} < \bar e$.
Therefore, the contribution of photometric error included in ${\rm SF}_{\rm obs}$ should be subtracted in order to unveil the intrinsic SF.

However, subtracting photometric error from ${\rm SF}_{\rm obs}$ is non-trivial. Many treatments have been proposed but are often imperfect in the sense that the inferred ``intrinsic'' SF has a biased amplitude or/and short-term slope \citep[see][for a review]{Kozlowski2016ApJ...826..118K}. 
In the following Sections~\ref{subsec:previous_sf} and \ref{subsect:p92_vs_s17}, we would briefly introduce three previous SF estimators and determine the best one among them. Note that two of the three previous SF estimators have been involved in \citet{Kozlowski2016ApJ...826..118K}. 
Since ${\rm SF}_{\rm true}$ for each realization is affected by the randomness of the simulation and so it does not completely agree with the input ${\rm SF}_{\rm input}$, we would compare the recovered intrinsic SF (hereafter ${\rm SF}_{\rm int}$) with ${\rm SF}_{\rm true}$ of each realization.

\subsection{P92 versus K16}\label{subsec:previous_sf}

\begin{figure*}
    \centering
    \includegraphics[width=0.32\linewidth]{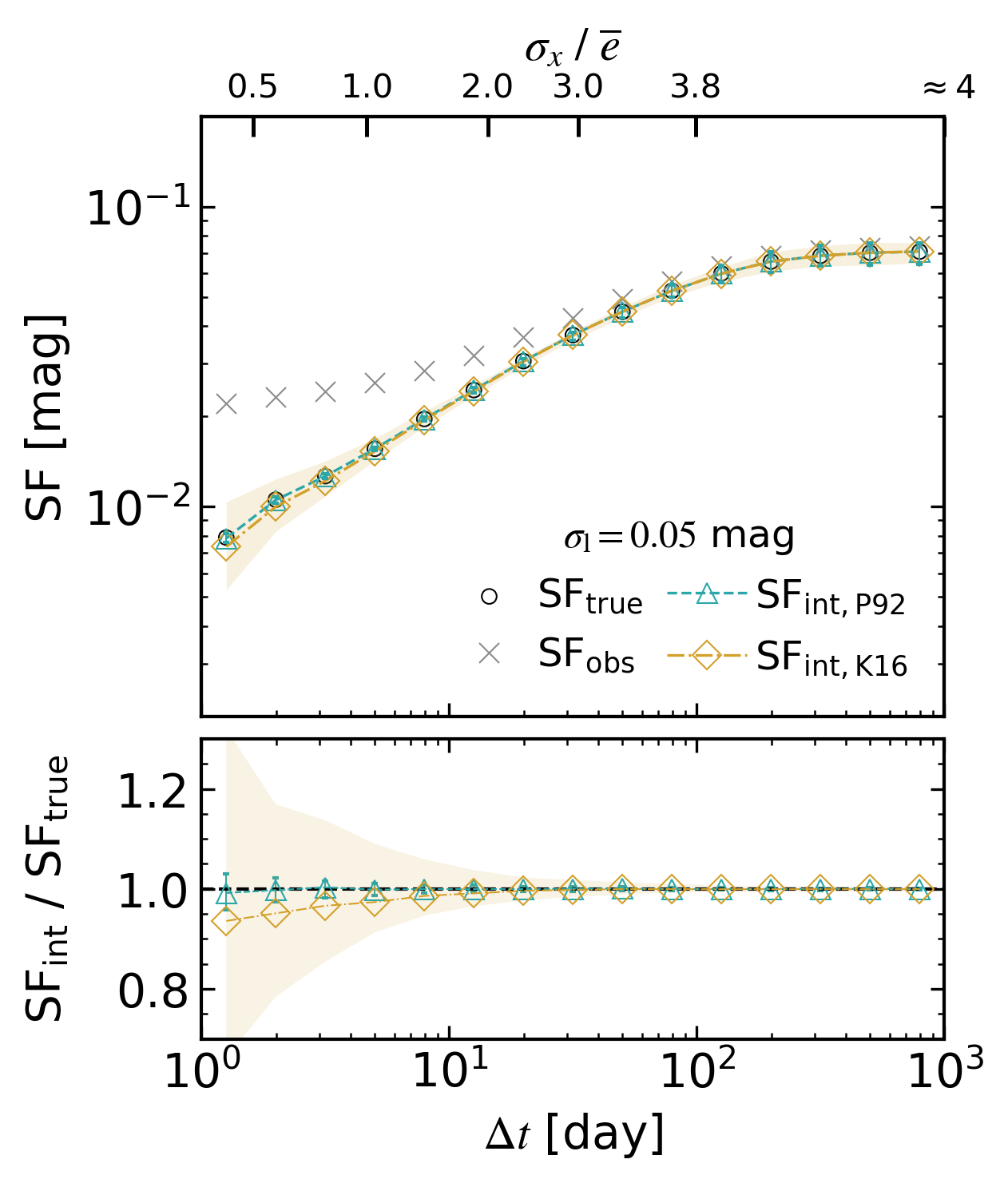} 
    \includegraphics[width=0.32\linewidth]{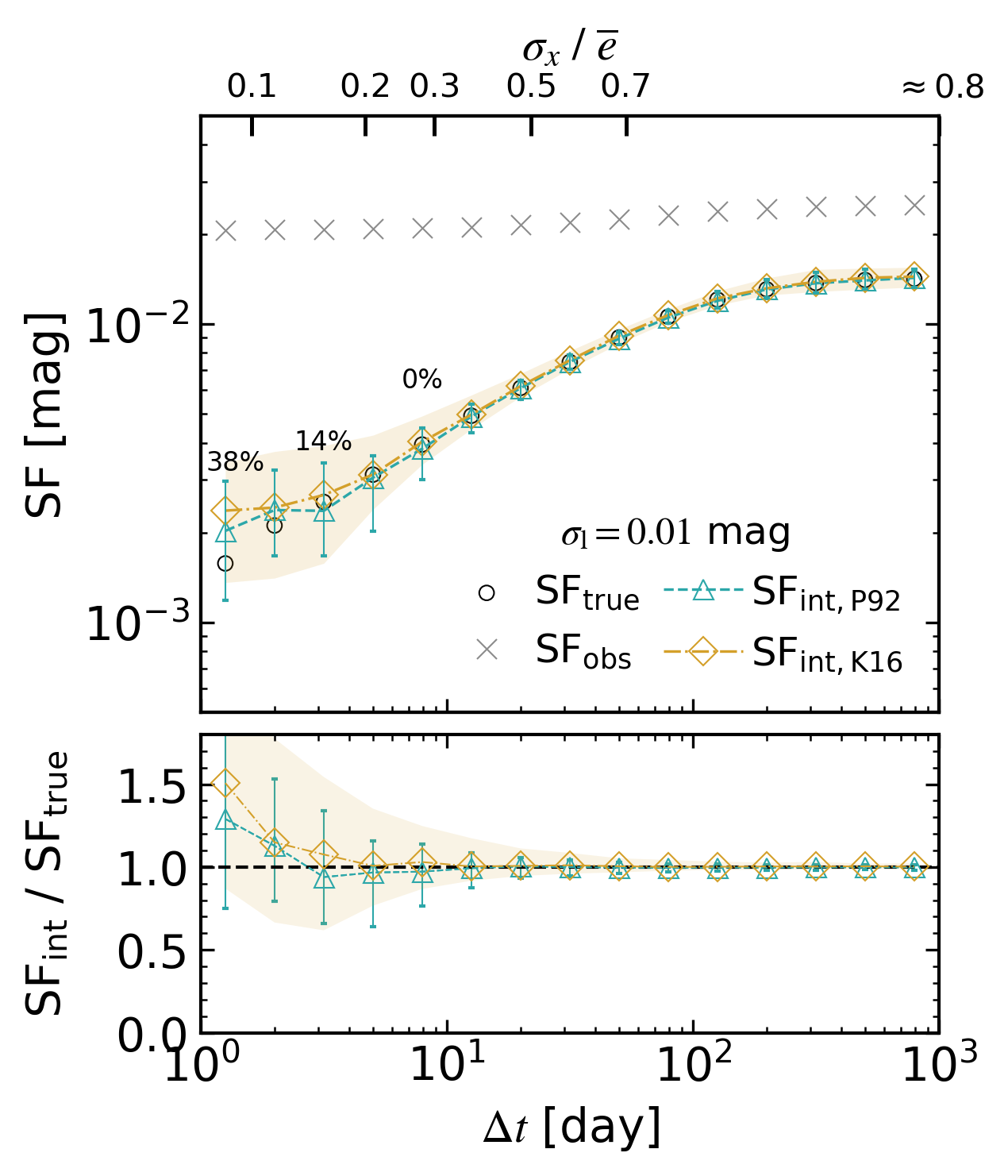} 
    \includegraphics[width=0.32\linewidth]{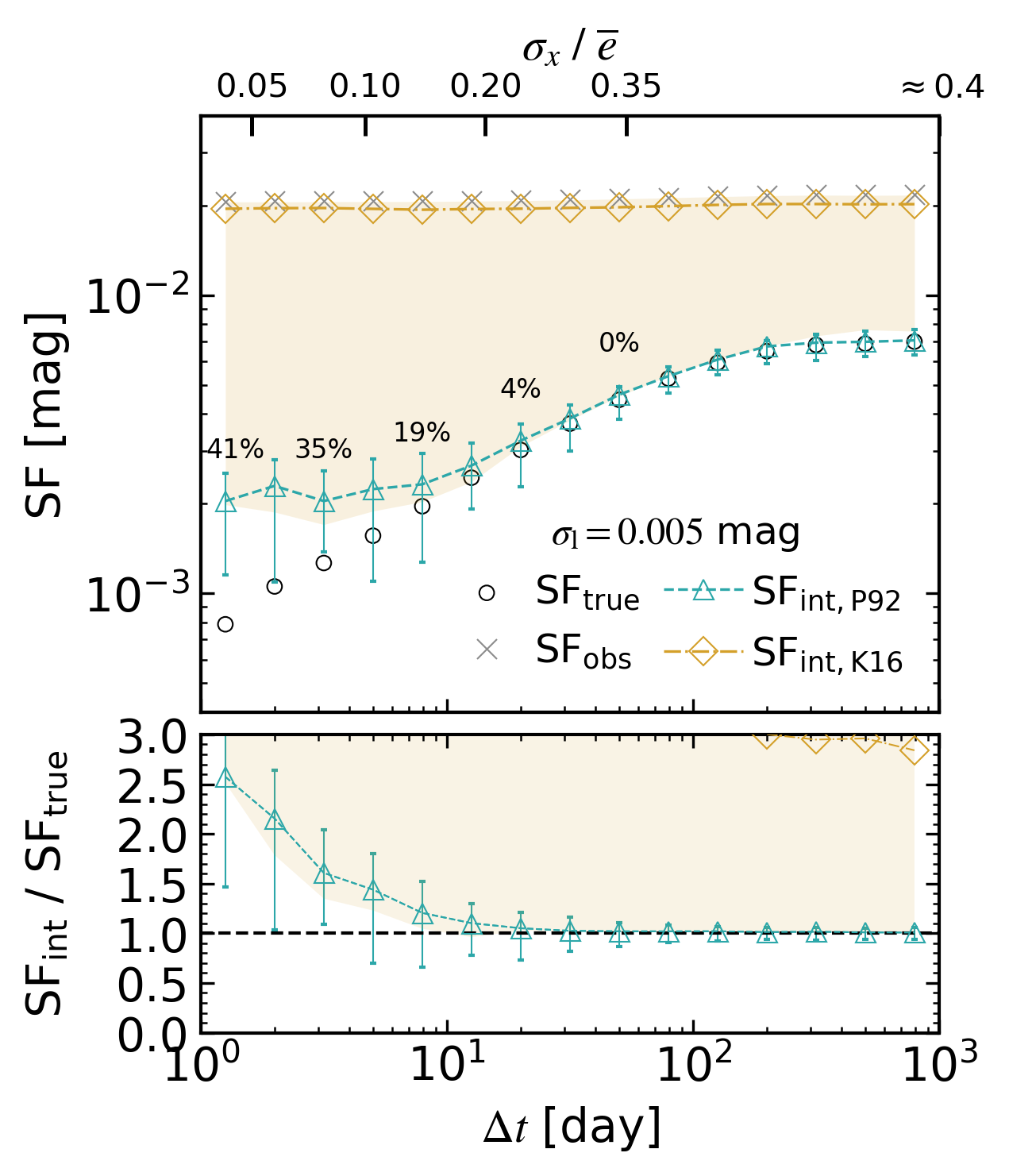}
    \caption{
    Illustration of the performance of the P92 and K16 estimators as decreasing $\sigma_{\rm l}$ from 0.05~mag (left column) to 0.01~mag (middle) and 0.005~mag (right), with fixed mean photometric error $\bar e \simeq 0.013$. 
    Based on 100 independent simulations, each column's top panel shows the median values of the true SF (${\rm SF}_{\rm true}$; circles), observed SF (${\rm SF}_{\rm obs}$; crosses), and intrinsic SFs retrieved from P92 (${\rm SF}_{\rm int,P92}$; triangles) and K16 (${\rm SF}_{\rm int,K16}$; diamonds). The bottom panel displays the corresponding ratios ${\rm SF}_{\rm int,P92}/{\rm SF}_{\rm true}$ (triangles) and ${\rm SF}_{\rm int,K16}/{\rm SF}_{\rm true}$ (diamonds).
    Vertical error bars and shaded region indicate the 16$-$84th percentile ranges over 100 realizations for P92 and K16, respectively.
    For P92, the numbers in small-$\Delta t$ bins give the fraction of realizations with ${\rm SF}_{\rm int,P92}^2 < 0$ due to noise over-subtraction.
    The top-axis of each column shows the ratio of the input timescale-dependent variation amplitude $\sigma_x(\Delta t)$ to $\bar e$, where $\sigma_x(\Delta t) = {\rm SF}_{\rm input}(\Delta t, \sigma_{\rm l}, \tau_{\rm d}, \beta) /\sqrt{2} = \sigma_{\rm l} \sqrt{1 - \exp{[-({\Delta t}/{\tau_{\rm d}})^\beta]}}$.
    }
    \label{fig:p92_vs_k16}
\end{figure*}

\begin{figure*}
    \centering
    \includegraphics[width=0.32\linewidth]{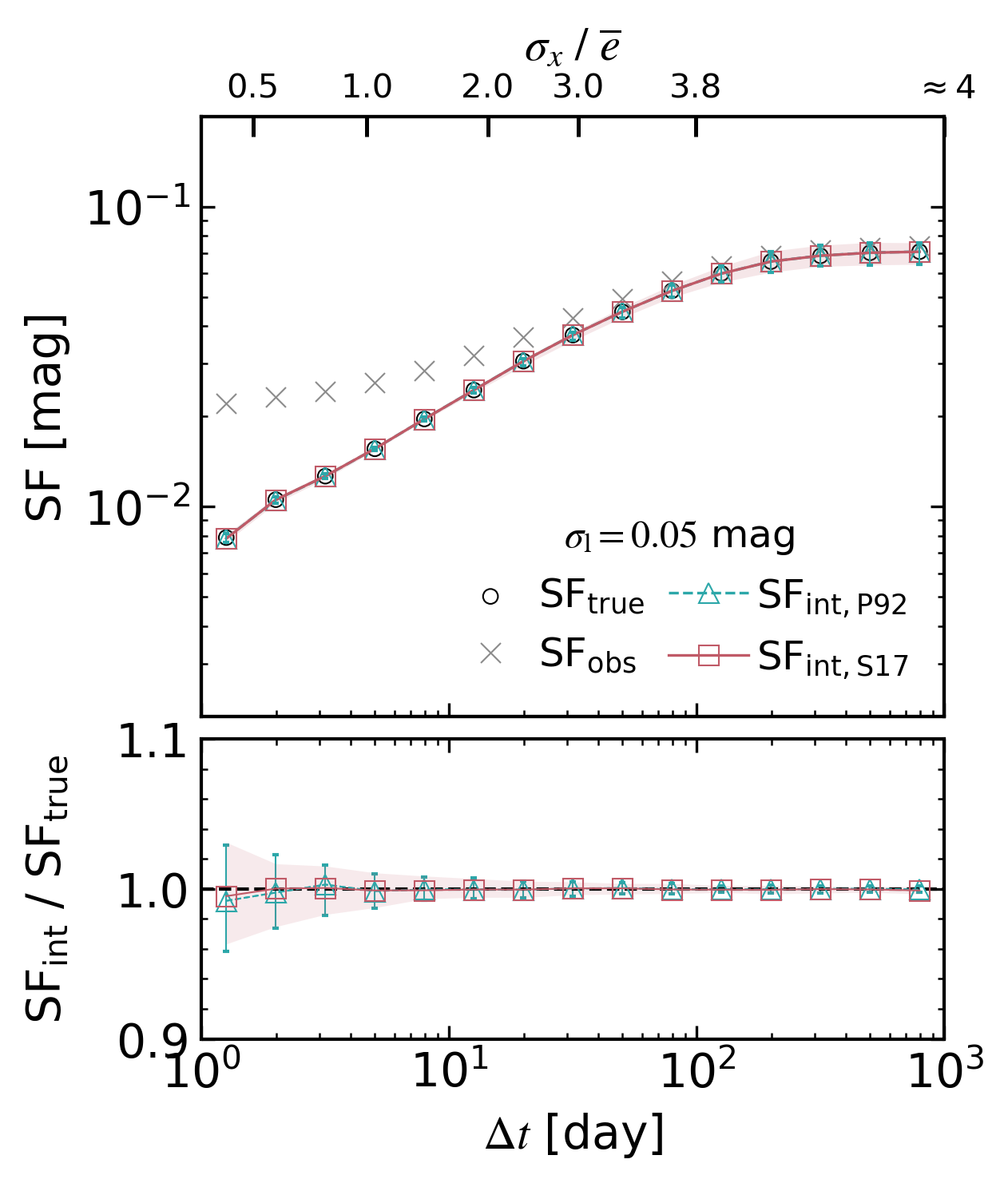} 
    \includegraphics[width=0.32\linewidth]{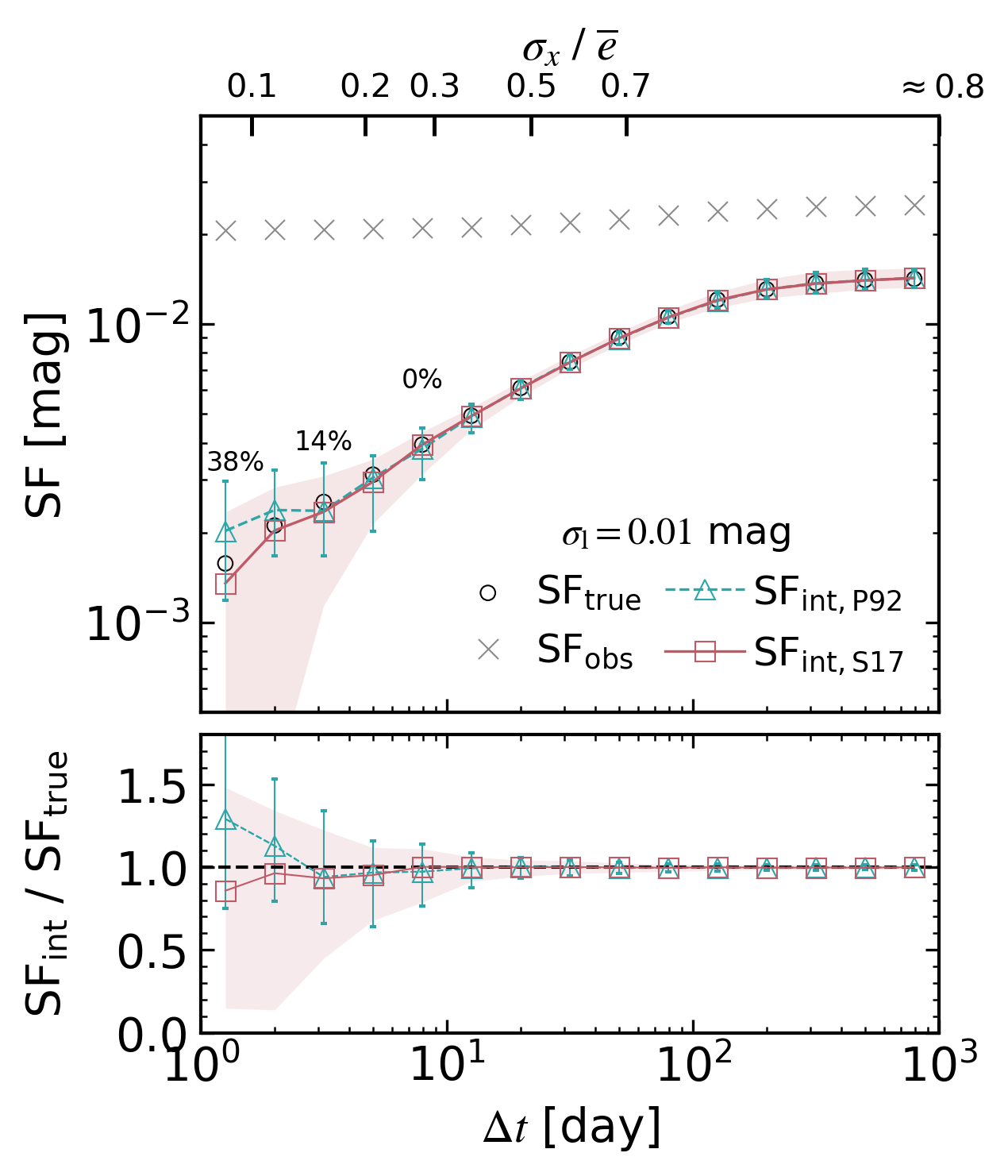} 
    \includegraphics[width=0.32\linewidth]{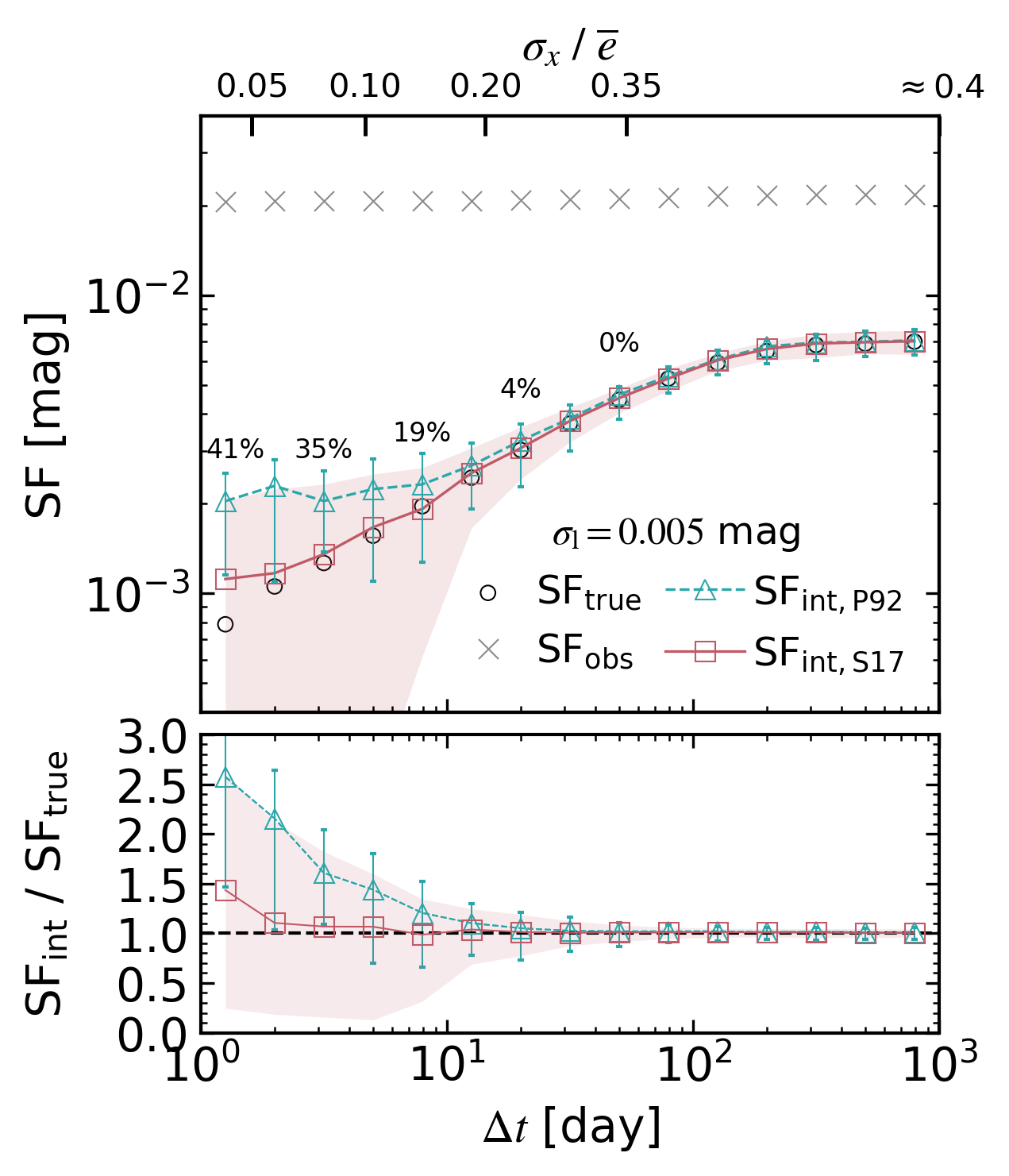} 
    \caption{Similar to Figure 4, but comparing the S17 estimator (squares) with the P92 estimator (triangles). The vertical error bars for P92 and the shaded region for S17 show the 16th--84th percentile ranges across 100 independent realizations. 
    For S17, the shaded region spans the 16th--84th percentiles of the 100 MAP estimates, showing the scatter across realizations rather than the posterior credible interval for an individual realization. 
    P92 realizations with ${\rm SF}_{\rm int,P92}^{2}<0$ are excluded before taking the square root and calculating the SF percentiles.}
    \label{fig:p92_vs_s17}
\end{figure*}

The first SF estimator was proposed by \citet[][hereafter P92]{Press1992}, who suggest subtracting the photometric errors using
\begin{align}\label{eq:p92}
{\rm SF}^2_{{\rm int,P92}}(\Delta t) &= { \frac{1}{N_{\rm pair}} \sum_{i=1}^{N_{\rm pair}} [(\Delta m_i')^2 - e_j^2 - e_k^2 ] }\nonumber\\
&= {\rm SF}^2_{{\rm obs}}(\Delta t) - \sigma^2_{{\rm n},j} - \sigma^2_{{\rm n},k},
\end{align}
where an $i$ corresponds to a $\{j,k\}$ pair and $\sigma^2_{{\rm n},j/k} \equiv \sum_{i=1}^{N_{\rm pair}} e_{j/k}^2 / N_{\rm pair}$ is the mean squared photometric errors at all relevant epochs $t_{j/k}$.
If the photometric errors are uncorrelated among different epochs, we would expect $\sigma^2_{{\rm n},j} = \sigma^2_{{\rm n},k}$. Note that $e_j$ should be estimated in prior from light curves of non-variable stars of the same magnitude.

The second SF estimator was offered by \citet[][hereafter K16]{Kozlowski2016ApJ...826..118K}, who fits a speculated input SF, e.g., Equation~(\ref{eq:sf_input}), plus a noise term, $\sigma_{\rm n}$, to the observed SF, that is,
\begin{equation}\label{eq:fit_k16}
{\rm SF}_{\rm obs}^2 (\Delta t) = {\rm SF}^2_{\rm input}(\Delta t \mid {\rm SF}_\infty,\tau_{\rm d},\beta) + 2\sigma_{\rm n}^2,
\end{equation}
where $\sigma_{\rm n}$ is mainly determined by the flat floor of observed SF at short timescales (cf. crosses in small-$\Delta t$ bins of Figure~\ref{fig:p92_vs_k16}). Once $\sigma_{\rm n}$ has been determined, the intrinsic SF is given by 
\begin{equation}\label{eq:sf_fit_k16}
{\rm SF}_{\rm int,K16}^2 (\Delta t) = {\rm SF}_{\rm obs}^2 (\Delta t) - 2\sigma_{\rm n}^2.
\end{equation}

For each simulated light curve (Section~\ref{sec:sim_lc}), we retrieve the intrinsic SF using the P92 and K16 estimators. To quantify their performance, we present the median and 16-84th percentile ranges of 100 SFs retrieved from 100 independent realizations (Figure~\ref{fig:p92_vs_k16}). Note that in the weak-variability regime and in small $\Delta t$ bins, both ${\rm SF}_{\rm int,P92}^2 (\Delta t)$ and ${\rm SF}_{\rm int,K16}^2 (\Delta t)$ may be negative in some realizations due to noise over-subtraction; these negative values are excluded before computing the median and percentiles. For example, in the middle and left columns of Figure~\ref{fig:p92_vs_k16}, the number in each small-$\Delta t$ bin give the fraction of realizations with ${\rm SF}_{\rm int,P92}^2 (\Delta t) < 0$ out of 100 realizations.

Figure~\ref{fig:p92_vs_k16} shows the performances of the P92 and K16 estimators on retrieving the intrinsic SF.
When $\sigma_{\rm l} > \bar e$, e.g., $\sigma_{\rm l} / \bar e \simeq 4$ in the left column of Figure~\ref{fig:p92_vs_k16}, both SF estimators work well though the K16 estimator results in larger scatters at short timescales. Once $\sigma_{\rm l} \simeq \bar e$, e.g., $\sigma_{\rm l} / \bar e \simeq 0.8$ in the middle column of Figure~\ref{fig:p92_vs_k16}, both SF estimators imply large scatters at short timescales and more importantly display a systematic overestimation of SF. This overestimation gets more prominent and appears on longer timescales with decreasing $\sigma_{\rm l}$. For example, when $\sigma_{\rm l} / \bar e \simeq 0.4$ in the right column of Figure~\ref{fig:p92_vs_k16}, the P92 estimator overestimates the SF at $\Delta t \lesssim 10$~d, while the K16 estimator is completely runaway over all timescales.

We therefore conclude that P92 outperforms K16 across a wide range of $\sigma_{\rm l}/\bar{e}$. This is not to say that P92 is unbiased: for $\sigma_{\rm l} \lesssim \bar{e}$ it frequently over-subtracts the photometric error, yielding negative ${\rm SF}_{\rm int,P92}^2 (\Delta t)$ in a large fraction of realizations (see numbers in the middle and right columns of Figure~\ref{fig:p92_vs_k16}). Because such realizations have no real square root and are often discarded, the surviving ones are biased upward--the origin of the overestimation noted above, which we revisit in Section~\ref{sec:p92_floor}.

\subsection{P92 versus S17}\label{subsect:p92_vs_s17}

The third SF estimator was introduced by \citet[][hereafter S17]{Suberlak2017}, who assess the intrinsic SF within each $\Delta t$-bin using a Bayesian likelihood method \citep[cf.][]{Almaini2000}.
This method assesses the intrinsic SF within any $\Delta t$-bin by maximizing the likelihood
\begin{align}\label{sf:s17_likelihood}
p(\{\Delta m_i'\}& \mid {\rm SF}_{\rm int,S17}, \mu, \{\epsilon_i\}) \nonumber \\
&= \prod_{i=1}^{N_{\rm pair}} \frac{1}{\sqrt{2\pi}\sigma_i} \exp\left[-\frac{(\Delta m_i'-\mu)^2}{2\sigma_i^2}\right],
\end{align}
where $\Delta m_i' = m_j' - m_k'$ is the magnitude difference, whose photometric error is $\epsilon_i = \sqrt{e^2_j + e^2_k}$, and
\begin{equation}\label{eq:s17_model}
\sigma_i^2 = {\rm SF}^2_{\rm int,S17} + \epsilon_i^2.
\end{equation}
The parameter $\mu$ is introduced to account for possible systematic photometric errors (e.g., the offset of zero point) between observing epochs.
Following \citet{Suberlak2017} we assume uniform priors for SF$_{\rm int,S17}$ and $\mu$, and adopt the maximum a posteriori (MAP) estimates for them. 
Here, we adopt prior ranges of ${\rm SF}_{\rm int,S17} \in [10^{-8},1]$~mag and $\mu \in [-1,1]$~mag.
The $1\sigma$ uncertainty of one parameter, e.g., SF$_{\rm int,S17}$, is given by the 16$-$84th percentile range of its one-dimensional posterior probability distribution after marginalizing over the other parameter, e.g., $\mu$.

Figure~\ref{fig:p92_vs_s17} demonstrates that the S17 estimator performs better than the P92 one. For $\sigma_{\rm l} > \bar{e}$ (left column), both recover the true SF accurately and with comparable scatter. Once $\sigma_{\rm l} \lesssim \bar{e}$, however, they diverge in character. 
At short timescales, the P92 estimator generally overestimates the intrinsic SF because it is subject to some realizations with ${\rm SF}_{\rm int,P92}^2 (\Delta t) < 0$, whereas the S17 estimator systematically performs well, albeit with large scatter around the intrinsic SF. In contrast to the P92 estimator, the S17 estimator always yields positive ${\rm SF}_{\rm int,S17}^2 (\Delta t)$ by enforcing non-negativity via the prior, rather than by discarding negative bins. With accurate photometric errors, it neither over-subtracts the signal nor loses realizations. However, individual S17 estimates can depart from the true SF when the intrinsic signal is weak. 
See Sections~\ref{sec:p92_floor} and \ref{sec:s17_optimal} for further discussions.

\section{Discussions}\label{sec:discussion}
\subsection{Toward an unbiased measurement of variability properties}

\subsubsection{Statistical origin of the P92 short-timescale floor}\label{sec:p92_floor}

Figures~\ref{fig:p92_vs_k16} and \ref{fig:p92_vs_s17} show that the P92 method systematically overestimates the intrinsic SF at short timescales when $\sigma_x(\Delta t) / \bar e\lesssim0.1$, producing a noise floor whose amplitude increases as the intrinsic variability becomes weaker relative to the photometric uncertainty. 
To understand the origin of this floor, we examine the statistical properties of ${\rm SF}^2_{\rm int,P92}$ in the weak-signal regime.

We consider the pure-noise limit, where the intrinsic variability is negligible.
For an individual pair of magnitude difference, we write $\Delta m'_i=\epsilon_i Z_i$ with $Z_i \sim \mathcal{N}(0,1)$, so that $(\Delta m'_i)^2-\epsilon_i^2=\epsilon_i^2(Z_i^2-1)$.
This noise-subtracted term can be positive or negative, taking negative values when $|Z_i|<1$, i.e., for $68.3\%$ of pairs. This is consistent with the negative fraction of intraday noise-subtracted pairs reported by \citet{Tang2024}. 
Assuming approximately uniform uncertainties with $\epsilon_i \simeq \epsilon \simeq \sqrt{2}\,\bar{e}$, the P92 estimator (Equation~\ref{eq:p92}) becomes
\begin{equation}\label{eq:sf2hat}
\begin{split}
{\rm SF}^2_{\rm int,P92} &= \frac{1}{N_{\rm pair}}\sum_{i=1}^{N_{\rm pair}}\bigl[(\Delta m'_i)^2-\epsilon^2_i\bigr] \\
&\simeq \frac{1}{N_{\rm pair}}\sum_{i=1}^{N_{\rm pair}}\bigl[(\Delta m'_i)^2-\epsilon^2\bigr] \\
&= \epsilon^2(U-1),
\end{split}
\end{equation}
where $U\equiv\sum_{i=1}^{N_{\rm pair}}Z_i^2 / {N_{\rm pair}} = \chi^2(N_{\rm pair}) / N_{\rm pair}$ and $\chi^2(N_{\rm pair})$ denotes a chi-squared random variable with $N_{\rm pair}$ degrees of freedom. 
Thus, $U$ has expectation unity, and ${\rm SF}^2_{\rm int,P92}$ has expectation zero.
For large $N_{\rm pair}$, both $U$ and ${\rm SF}^2_{\rm int,P92}$ are approximately normally distributed. Specifically, $U-1$ is approximately distributed as $\sqrt{2/N_{\rm pair}}\,Z'$, where $Z'\sim\mathcal{N}(0,1)$ and $\sqrt{2/N_{\rm pair}}$ is the standard deviation of $U=\chi^2(N_{\rm pair})/N_{\rm pair}$.
Hence, in about half of the realizations, ${\rm SF}^2_{\rm int,P92}$ is negative. 
Since the SF amplitude is generally computed as ${\rm SF}_{\rm int,P92}=\sqrt{{\rm SF}^2_{\rm int,P92}}$, negative ${\rm SF}^2_{\rm int,P92}$ cannot be converted into physical SF amplitudes. 
Therefore, removing these negative realizations biases the distribution of measurable SF values upward, producing an artificial non-zero floor. Its characteristic amplitude is given by
\begin{equation}\label{eq:sffloor_def}
{\rm SF}_{\rm floor}=\epsilon\,\underset{U > 1}{\mathrm{med}}\left(\sqrt{U-1}\right),
\end{equation}
where the median is taken over realizations with $U>1$, i.e., positive ${\rm SF}^2_{\rm int,P92}$.
The condition $U>1$ corresponds to $Z'>0$, which follows a half-normal distribution. Since the square root is a monotonic, ${\rm med}(\sqrt{Z'})=\sqrt{{\rm med}(Z')}$. For the half-normal distribution, $2\Phi[{\rm med}(Z')]-1=0.5$,
where $\Phi$ is the cumulative distribution function of the standard normal distribution, yielding
${\rm med}(Z')=\Phi^{-1}(0.75)=0.674$~\footnote{For the half-normal distribution of $Z'$ conditional on $Z'>0$, the mean of $\sqrt{Z'}$ is approximately 0.822, close to its median, $\sqrt{0.674}\simeq0.821$. Thus, using the mean instead of the median has a negligible effect on the floor predicted under this approximation.}.
Consequently, we have
\begin{align}\label{eq:sffloor}
{\rm SF}_{\rm floor}&\approx\sqrt{2}\,\bar e\left(\frac{2}{N_{\rm pair}}\right)^{1/4}\sqrt{0.674}\nonumber\\
&\approx0.821\,\sqrt{2}\,\bar e\left(\frac{2}{N_{\rm pair}}\right)^{1/4}.
\end{align}

According to Equation~(\ref{eq:sffloor}), this floor is determined solely by the photometric uncertainty and the number of pairs, rather than by the intrinsic variability signal. Moreover, given ${\rm SF}_{\rm floor}\propto\bar e\,N_{\rm pair}^{-1/4}$, its dependence on the number of pairs is weak, so the floor remains nearly constant across short timescales, producing the horizontal feature seen in Figures~\ref{fig:p92_vs_k16} and \ref{fig:p92_vs_s17}. For our shortest bin, with $\bar e\simeq0.013$~mag and $N_{\rm pair}\approx4\times10^4$, Equation~(\ref{eq:sffloor}) gives ${\rm SF}_{\rm floor}\approx1.3\times10^{-3}$~mag, marginally below the measured floor at $\Delta t\lesssim10$~d. 
The analytic value is expected to be somewhat lower than the measured one, because it assumes zero intrinsic signal, identical uncertainties, and independent pairs. In real light curves, shared epochs correlate the pairs, reducing the effective number of independent pairs below $N_{\rm pair}$ and thus raising the floor.

\begin{figure*}
    \centering
    \includegraphics[width=0.32\linewidth]{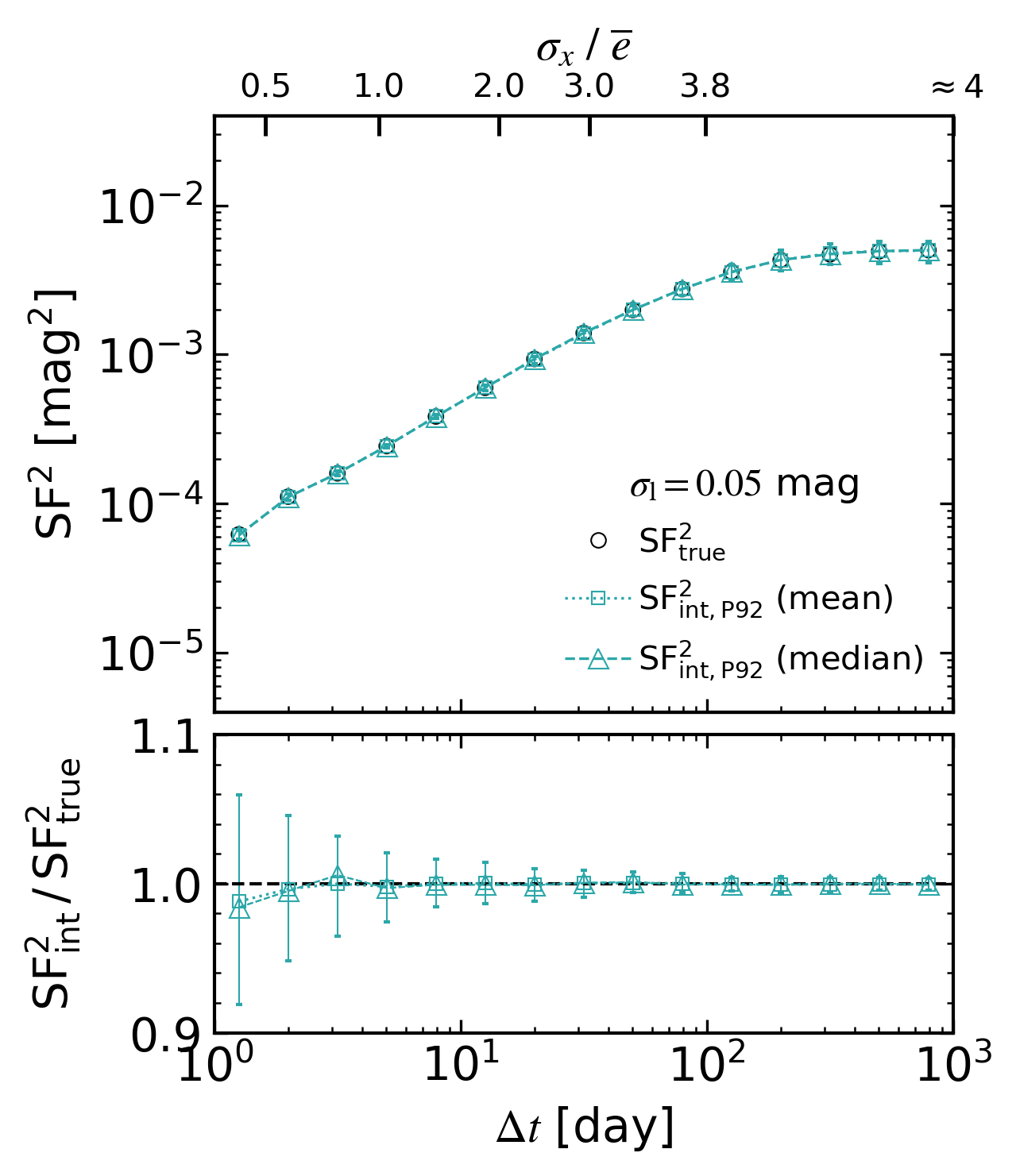} 
    \includegraphics[width=0.32\linewidth]{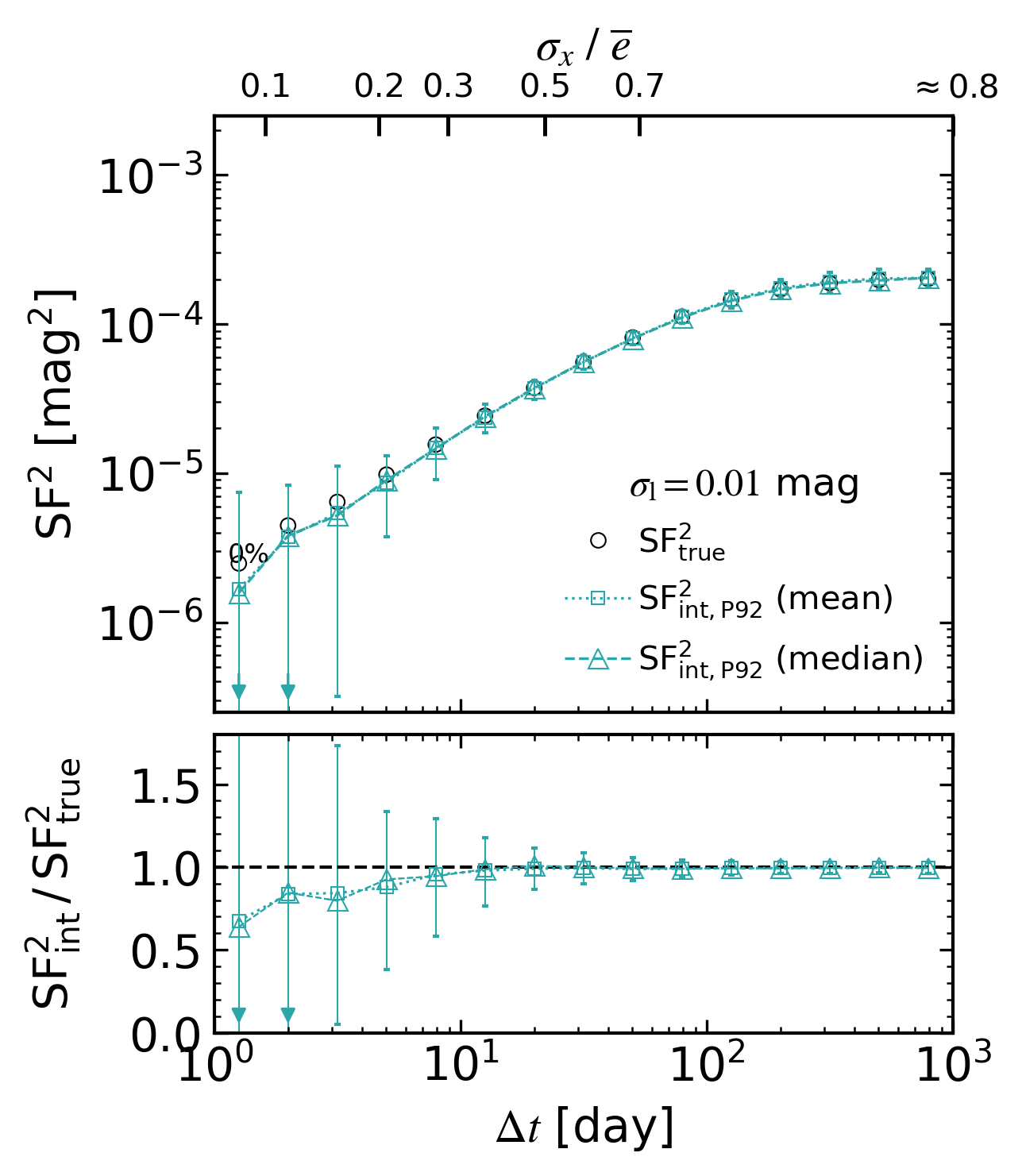} 
    \includegraphics[width=0.32\linewidth]{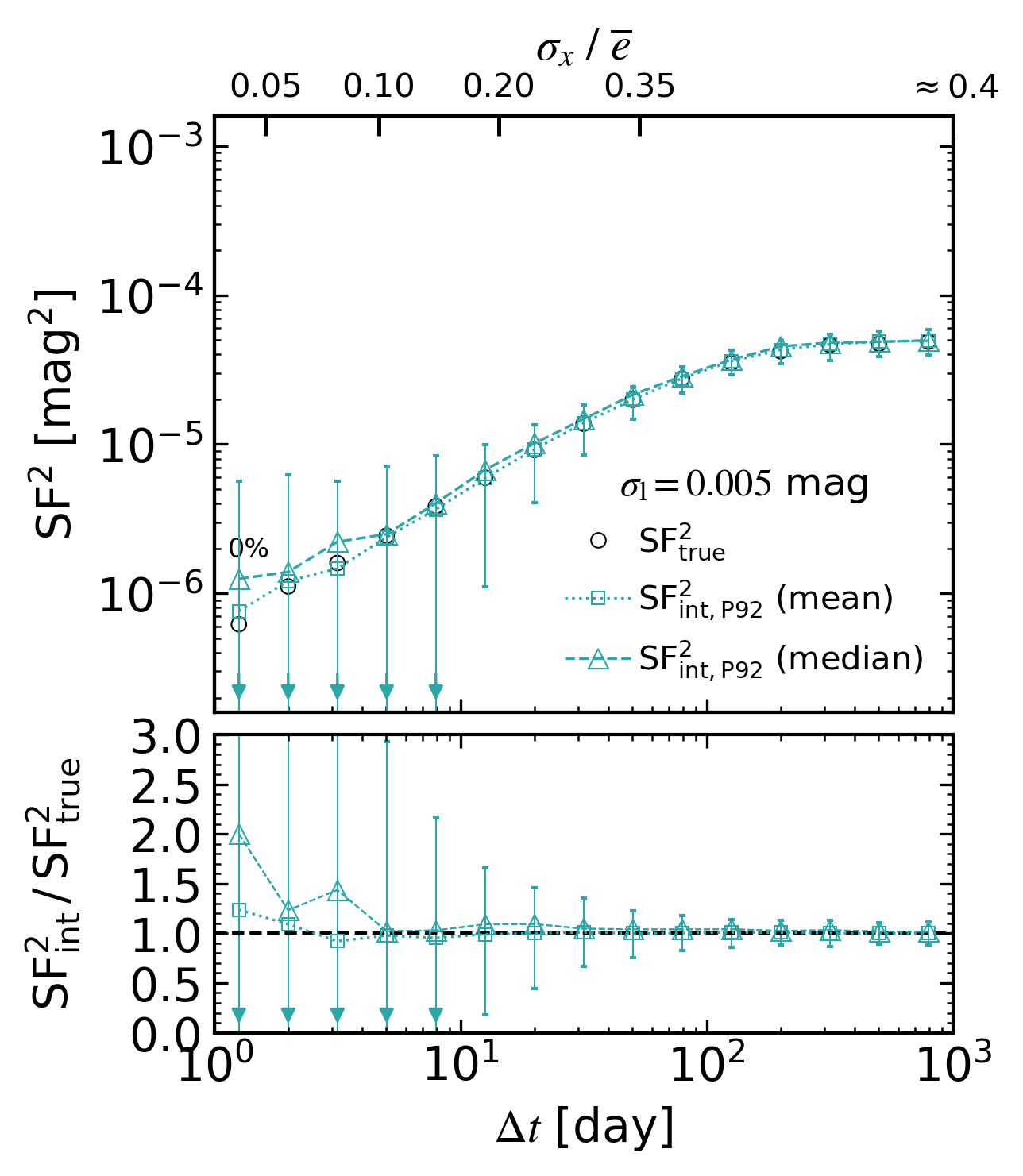} 
    \caption{{P92 estimates of the intrinsic $\mathrm{SF}^2$, retaining negative values after noise subtraction. Black circles show the median $\mathrm{SF}_{\rm true}^2$ from the ideal light curves, while triangles (squares) and error bars show the median (mean) and 16th--84th percentile ranges of the P92 estimates across 100 realizations. 
    The lower panels show the corresponding ratios $\mathrm{SF}_{\rm int,~P92}^2/\mathrm{SF}_{\rm true}^2$.
    In the weak-signal regime, some 16th-percentile bounds fall below zero and the downward arrows indicate 16th-percentile bounds below zero, which lie outside the displayed axis limits.
    }}
    \label{fig:p92_sf2}
\end{figure*}

Given this floor, a subsequent question concerns the timescales over which the P92 estimator can provide reliable estimates. This can be assessed by comparing ${\rm SF}_{\rm floor}$ with the input SF, i.e., ${\rm SF}_{\rm floor} \sim {\rm SF}_{\rm input}(\Delta t) = \sqrt{2} \sigma_x(\Delta t)$, which yields ${\sigma_x(\Delta t)}/{\bar e} \sim 0.82~\left({2}/{N_{\rm pair}}\right)^{1/4}$ so that ${\sigma_x(\Delta t)}/{\bar e} \sim 0.1$ for $N_{\rm pair} \sim 10^4$, consistent with our simulation results that the floor appears at  $\sigma_x(\Delta t) / \bar e\lesssim0.1$ (Figures~\ref{fig:p92_vs_k16} and~\ref{fig:p92_vs_s17}).
In practice, for each $\Delta t$ bin, based on its $\bar e$ and $N_{\rm pair}$, Equation~(\ref{eq:sffloor}) provides a useful criterion for checking the reliability of measured ${\rm SF}_{\rm int,P92}(\Delta t)$ by comparing it to ${\rm SF}_{\rm floor}(\Delta t)$ estimated in that bin: if ${\rm SF}_{\rm floor}(\Delta t) \gtrsim {\rm SF}_{\rm int,P92}(\Delta t)$, then the measured ${\rm SF}_{\rm int,P92}(\Delta t)$ should be treated with caution.

Although increasing $N_{\rm pair}$ can help lower the floor, doing so is not straightforward in practice: a tenfold reduction in ${\rm SF}_{\rm floor}$ requires a $10^4$-fold increase in $N_{\rm pair}$. Given the finite baselines and sparse sampling of individual AGNs, even ensemble averaging over many AGNs offers only marginal improvement. Thus, a floor is commonly seen at short timescales in real dataset \citep[e.g.,][]{DeCicco2022,Tang2024}.

In short, the floor discussed above arises from excluding negative estimates when calculating the SF amplitude. 
In the weak-variability regime, presenting only the square roots of positive P92 estimates can therefore give a misleading impression of a nonzero intrinsic signal. 
Figure~\ref{fig:p92_sf2} shows that retaining negative values when calculating the median ${\rm SF}^2$ (or the mean, as in \citealt{Caplar2017}) avoids this selection bias, with both remaining close to the true value across 100 ideally sampled realizations, although the 16th--84th percentile ranges are broad when the signal is weak. However, real observations generally have sparser and less regular sampling, and retaining negative values alone does not guarantee unbiased recovery. Weak variability can still be poorly constrained by noisy data, as discussed in Section~\ref{sec:s17_optimal} and Section~\ref{subsec:sampling}.

\subsubsection{The noise--signal degeneracy in the K16 fit}\label{sec:k16_degeneracy}

A natural alternative to pairwise noise subtraction (i.e., the P92 estimator) is to jointly model the noise with the intrinsic signal (cf. the K16 estimator). Following \citet{Kozlowski2016ApJ...826..118K}, we fit the four-parameter model of Equation~(\ref{eq:fit_k16}) to the observed SF via least squares, allowing both the common noise level $\sigma_{\rm n}$ and the signal parameters $({\rm SF}_\infty,\tau_{\rm d},\beta)$ to vary freely. This does not rely on the measured photometric errors, but introduces a degeneracy between $\sigma_{\rm n}$ and the weak short-timescale signal, particularly the parameter $\beta$.

At short timescales, where $\sigma_x\ll\sigma_{\rm n}$, the fit determines $\sigma_{\rm n}$ from ${\rm SF}_{\rm obs}\approx\sqrt2\,\sigma_{\rm n}$. At long timescales, ${\rm SF}_{\rm obs}$ saturates at $\approx\sqrt{2\sigma_{\rm n}^2+{\rm SF}_\infty^2}$; subtracting the $\sigma_{\rm n}$ derived at short timescales yields ${\rm SF}_\infty$. 
Between these two regimes, ${\rm SF}_{\rm obs}$ rises from $\approx\sqrt2\,\sigma_{\rm n}$ to $\approx\sqrt{2\sigma_{\rm n}^2+{\rm SF}_\infty^2}$, and this transition constrains the remaining two parameters, i.e., the slope $\beta$ and the characteristic timescale $\tau_{\rm d}$. 
Thus, as illustrated in Figure~\ref{fig:p92_vs_k16}, the K16 estimator performs well only when the intrinsic signal is not much smaller than the noise, i.e., ${\rm SF}_\infty \gtrsim \sigma_{\rm n}$ (or equivalently $\sigma_x(\Delta \gg \tau_{\rm d}) \gtrsim \bar e$). 
If the short- and long-timescale plateaus are too close to be resolved, $\sigma_{\rm n}$ and ${\rm SF}_\infty$ cannot be separated, so $\beta$ and $\tau_{\rm d}$ are unconstrained (cf. the right column of Figure~\ref{fig:p92_vs_k16}).

The K16 estimator also suffers from another limitation: Equation~(\ref{eq:fit_k16}) assumes a single power law with one break, i.e., a single characteristic timescale $\tau_{\rm d}$. If the intrinsic AGN SF has more than one break (e.g., different slopes at short and long timescales), a single power law cannot fit both. Forced to describe two breaks with one, the fitted model matches neither slope, biasing $\tau_{\rm d}$ and the slope toward intermediate values and absorbing the residual into $\sigma_{\rm n}$. Consequently, ${\rm SF}_{\rm int,K16}$ becomes distorted even when the intrinsic signal is strong and the photometric errors are reliable, because the assumed functional form does not match the true SF.

In sum, the K16 estimator therefore could fail in two distinct ways. When the intrinsic signal is weak, $\sigma_{\rm n}$ and the intrinsic signal become degenerate.
When the intrinsic SF has more than one break, a single power law misrepresents the true shape, biasing the recovered slope and $\tau_{\rm d}$ even for a strong intrinsic signal. 

\subsubsection{S17 as a near-optimal estimator}\label{sec:s17_optimal}

Both P92 and S17 are non-parametric: each estimates ${\rm SF}_{\rm int}$ independently in every $\Delta t$ bin, without assuming a form for the intrinsic SF. Instead, K16 is parametric, fitting the whole observed SF with a four-parameter model. The three also differ in how they treat the noise: P92 subtracts it pair by pair, K16 fits it as a free parameter, and S17 incorporates the photometric errors directly into the likelihood. 

For each $\Delta t$ bin S17 models the magnitude differences as Gaussian samples with variance $\sigma_i^2={\rm SF}_{\rm int,S17}^2+\epsilon_i^2$, where the per-pair uncertainty $\epsilon_i$ is a known input, and infers ${\rm SF}_{\rm int,S17}$ from the posterior by MCMC (Equations~\ref{sf:s17_likelihood} and~\ref{eq:s17_model}). 
Because the photometric errors enter the likelihood as known inputs, S17 does not require discarding negative variance estimates.
However, its non-negativity constraint can still affect the MAP estimates in the weak-signal regime. As the signal weakens, the posterior becomes less informative, and individual estimates can show substantial scatter and departures from the true SF (cf. the shaded region in Figure~\ref{fig:p92_vs_s17}).

Thus, how well can the intrinsic SF be constrained in the weak-signal regime? For a bin of $N_{\rm pair}$ independent pairs with $\Delta m_i'\sim\mathcal{N}(\mu,\sigma_i^2)$ and $\sigma_i^2={\rm SF}_{\rm int,S17}^2+\epsilon_i^2$, the Cram\'er--Rao bound gives the minimum uncertainty on the intrinsic SF,
\begin{equation}
\delta{\rm SF}_{\rm int,S17}
\ge
\frac{1}{\sqrt{2N_{\rm pair}}}
\frac{{\rm SF}_{\rm int,S17}^2+\epsilon^2}{{\rm SF}_{\rm int,S17}},
\label{eq:s17_cr}
\end{equation}
the smallest error any unbiased estimator can achieve from the data. Requiring a $n\sigma$ detection, ${\rm SF}_{\rm int,S17}\ge n\,\delta{\rm SF}_{\rm int,S17}$, the weak-signal limit (${\rm SF}_{\rm int,S17}^2\ll\epsilon^2$, $\epsilon^2\simeq2\bar e^2$, ${\rm SF}_{\rm int,S17}=\sqrt2\,\sigma_x$) gives
\begin{equation}
\frac{\sigma_x}{\bar e}
\gtrsim
\sqrt{n}\left(2N_{\rm pair}\right)^{-1/4},
\label{eq:s17_thresh}
\end{equation}
which for our ideal sampling corresponds to $\sigma_x/\bar e\gtrsim0.07$ for a $1\sigma$ detection ($n=1$) and $0.10$ for $2\sigma$ ($n=2$). This is a detection limit set by Fisher information of the $\Delta m'$ data alone: below it the intrinsic variability is difficult to separate from the noise regardless of the estimator. 
Its value depends only on the photometric error and the number of pairs, so it sets which sources yield a real measurement and, inverted, how many pairs, and hence what baseline and cadence, are needed to reach a desired $\sigma_x/\bar e$. 
The limit is visible in Figure~\ref{fig:p92_vs_s17}: near $\sigma_x/\bar e\approx0.1$, the S17 posterior broadens sharply, and the S17 and P92 estimators begin to diverge. Below this threshold, their responses differ fundamentally: S17 translates the lost information into growing uncertainties in ${\rm SF}_{\rm int,S17}$, whereas P92 returns biased estimates pinned to its floor.

Finally, like P92, S17 assumes that the input photometric errors are accurate. When this condition is satisfied and sufficient informative pairs are available, S17 generally recovers the true SF across realizations, although individual MAP estimates can deviate substantially in the weak-signal regime. The Cram\'er--Rao bound provides an idealized precision limit set by the photometric uncertainties and the number of independent pairs (Equation~\ref{eq:s17_thresh}). The advantage of S17 is therefore not higher sensitivity, but a more faithful representation of the intrinsic variability and its uncertainty.

\begin{figure}
    \centering
    \includegraphics[width=0.9\linewidth]{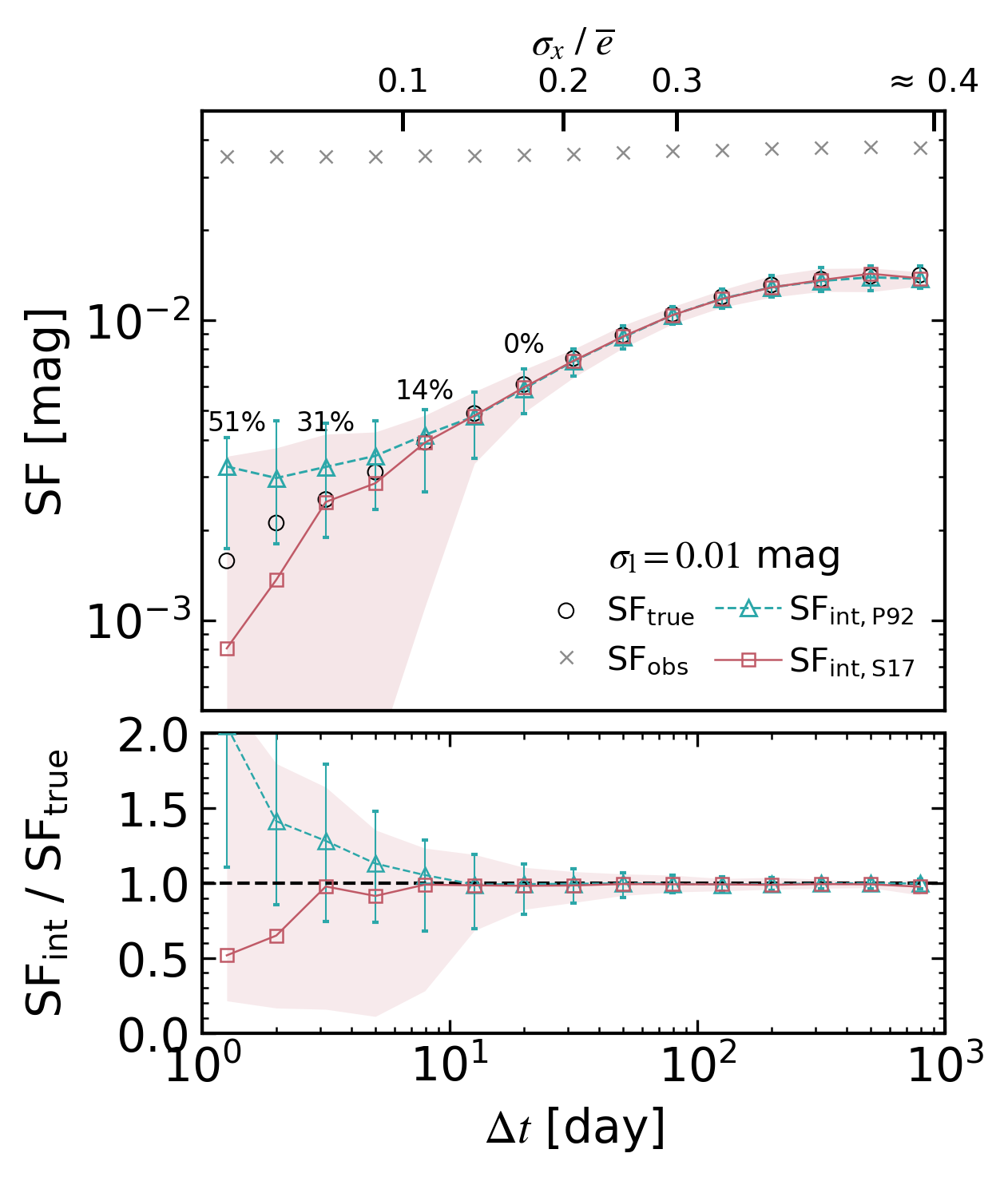}
    \caption{Similar to the middle panel of Figure~\ref{fig:p92_vs_s17}, but for simulated light curves with a mean magnitude of 20 mag. 
    }
    \label{fig:sf_20mag}
\end{figure}

\subsection{Recovery of Intrinsic Variability for Fainter AGNs}
\label{subsec:photometric_error}

In Section~\ref{sec:measure_sf}, we considered a fiducial mean magnitude of $\bar m=17$ mag, for which the adopted photometric-error model gives a mean single-epoch uncertainty of $\bar e\simeq0.013$ mag. However, the median $r$-band magnitude of quasars in the SDSS Stripe 82 sample is close to 20 mag \citep{MacLeod2010ApJ...721.1014M}; we therefore also consider a fainter case with $\bar m=20$ mag. According to the same photometric-error model (Section~\ref{subsec:real_lc}; Figure~\ref{fig:e_sigma_e_mag}), this gives $\bar e\simeq0.025$ mag. For fixed intrinsic variability, i.e., with $\sigma_{\rm l}=0.01$ mag, the ratio $\sigma_{\rm l}/\bar e$ thus decreases from $\simeq0.8$ to $\simeq0.4$.

Figure~\ref{fig:sf_20mag} shows the resulting SF estimates for the case of $\bar m=20$ mag. At short timescales, where $\sigma_x(\Delta t)/\bar e\lesssim0.1$, the P92 estimates exhibit the amplitude floor described by Equation~(\ref{eq:sffloor}). Negative ${\rm SF}^2$ estimates occur in the first several timescale bins but are absent at $\Delta t\gtrsim20$ days.
The S17 estimates also depart from the true SF in this low-signal regime and show larger scatter across realizations as the relative signal weakens, consistent with Equation~(\ref{eq:s17_thresh}).

For fixed $\sigma_{\rm l}=0.01$ mag, increasing the photometric error from the $\bar m=17$ mag case (the middle panel of Figure~\ref{fig:p92_vs_s17}) to the $\bar m=20$ mag case (Figure~\ref{fig:sf_20mag}) makes the P92 floor apparent at longer timescales.
In both cases, the P92 floor becomes important when $\sigma_x(\Delta t)/\bar e\lesssim0.1$, corresponding to $\Delta t\lesssim2$ days at $\bar m=17$ mag and $\Delta t\lesssim7$ days at $\bar m=20$ mag. For S17, large scatter across realizations is seen up to $\Delta t\sim10$ days at $\bar m=20$ mag, compared with a few days at $\bar m=17$ mag.
Interestingly, the $\bar m=20$ mag, $\sigma_{\rm l}=0.01$ mag case (Figure~\ref{fig:sf_20mag}) has nearly the same $\sigma_{\rm l}/\bar e$ as the $\bar m=17$ mag, $\sigma_{\rm l}=0.005$ mag case (the right panel of Figure~\ref{fig:p92_vs_s17}). Both exhibit a dominant P92 floor and substantial S17 scatter on similar timescales, shorter than $\Delta t\sim10$ days. Together, these two comparisons show that increasing the photometric error or reducing the intrinsic variation amplitude leads to similar limitations on SF recovery, consistent with Equations~(\ref{eq:sffloor}) and~(\ref{eq:s17_thresh}).

\begin{figure*}
    \centering
    \includegraphics[width=0.98\linewidth]{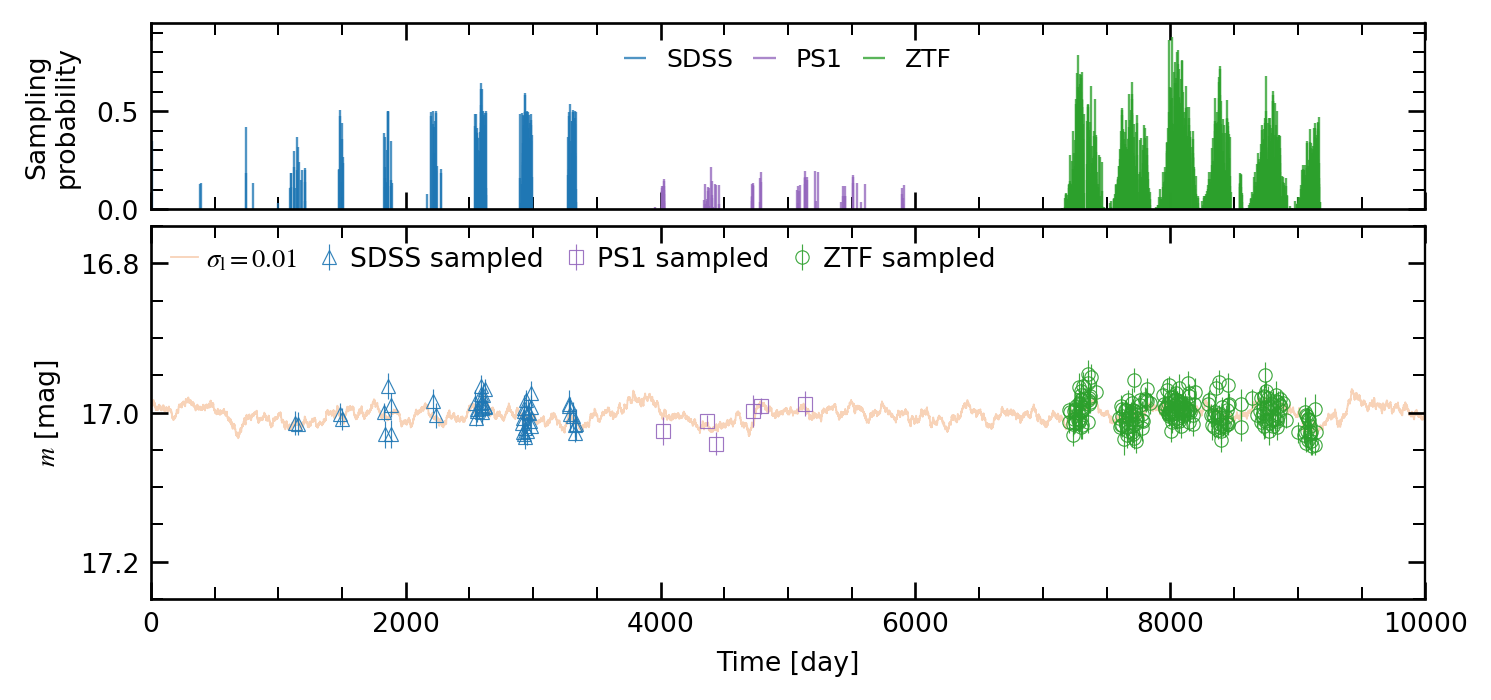}
    \caption{Illustration of a simulated light curve with $\sigma_{\rm l}=0.01$~mag sampled according to an empirical observing window. The upper panel shows the sampling probabilities constructed by combining SDSS \citep{York2000}, PS1 \citep{Chambers2016}, and ZTF \citep{Bellm2019} observations of Stripe 82 AGNs selected from \citet{MacLeod2010ApJ...721.1014M}. The lower panel shows the simulated ideal light curve (solid curve) and the simulated measurements retained according to the above probabilities: SDSS (triangles), PS1 (squares), and ZTF (circles).}
    \label{fig:sampling-lc}
\end{figure*}

\subsection{Recovery of Intrinsic Variability under Realistic Survey Sampling}\label{subsec:sampling}

Equation~\eqref{eq:s17_thresh} shows that the signal-to-noise ratio required to retrieve an unbiased intrinsic SF depends on the number of independent pairs in each timescale bin.
Real surveys generally have sparser and less regular sampling than our ideal simulations, resulting in fewer pairs.  We therefore test whether the intrinsic SF can still be reliably recovered under realistic survey sampling.

\citet{MacLeod2010ApJ...721.1014M} provided light curves for 9,258 quasars in the SDSS Stripe 82 region, based on SDSS observations spanning nearly a decade. Since then, time-domain surveys such as PS1 and ZTF have provided additional observations of quasars in this region, extending the baseline for studying their variability. 
Based on the quasar catalog of \citet{MacLeod2010ApJ...721.1014M}, we combine the SDSS, PS1, and ZTF $r$-band observations to construct an empirical sampling window (hereafter, the Stripe 82 cadence). 
For each quasar, we assign its observation epochs to a common time grid with a 0.5-day cadence, matching that of the simulated light curves in Section~\ref{sec:sim_lc}. 
If a quasar has multiple observations in one time bin, we count it only once. We define the sampling probability at grid time $t$ as $P(t)=\frac{1}{N_{\rm src}}\sum_{s=1}^{N_{\rm src}} I_s(t)$, where $N_{\rm src}$ is the number of quasars used to construct the window, and $I_s(t)=1$ if quasar $s$ was observed in the bin centered on $t$, and $I_s(t)=0$ otherwise. The upper panel of Figure~\ref{fig:sampling-lc} shows the sampling probabilities as a function of time for SDSS, PS1, and ZTF. The resulting window spans $\simeq9{,}200$ days ($\simeq25$ years), and the sum of the sampling probabilities corresponds to an expected $\simeq310$ sampled epochs per simulated light curve.

We apply this window to each of the 100 simulated ideal light curves by independently retaining each simulated epoch with probability $P(t)$. We use the same intrinsic light curves and photometric errors as in Section~\ref{sec:measure_sf}; only the observing cadence changes. 
The lower panel of Figure~\ref{fig:sampling-lc} shows an example from the 100 realizations: the $\sigma_{\rm l}=0.01$ mag simulated ideal light curve from Figure~\ref{fig:ideal-lc-sf}, together with the measurements retained by the sampling window.
Note that we here assume that all surveys have the same photometric-error properties and that there are no survey-dependent zero-point offsets. In real data, however, these differences need to be calibrated when combining observations from different surveys, for example, using non-variable stars \citep{pytics}.

\begin{figure*}
     \centering
    \includegraphics[width=0.32\linewidth]{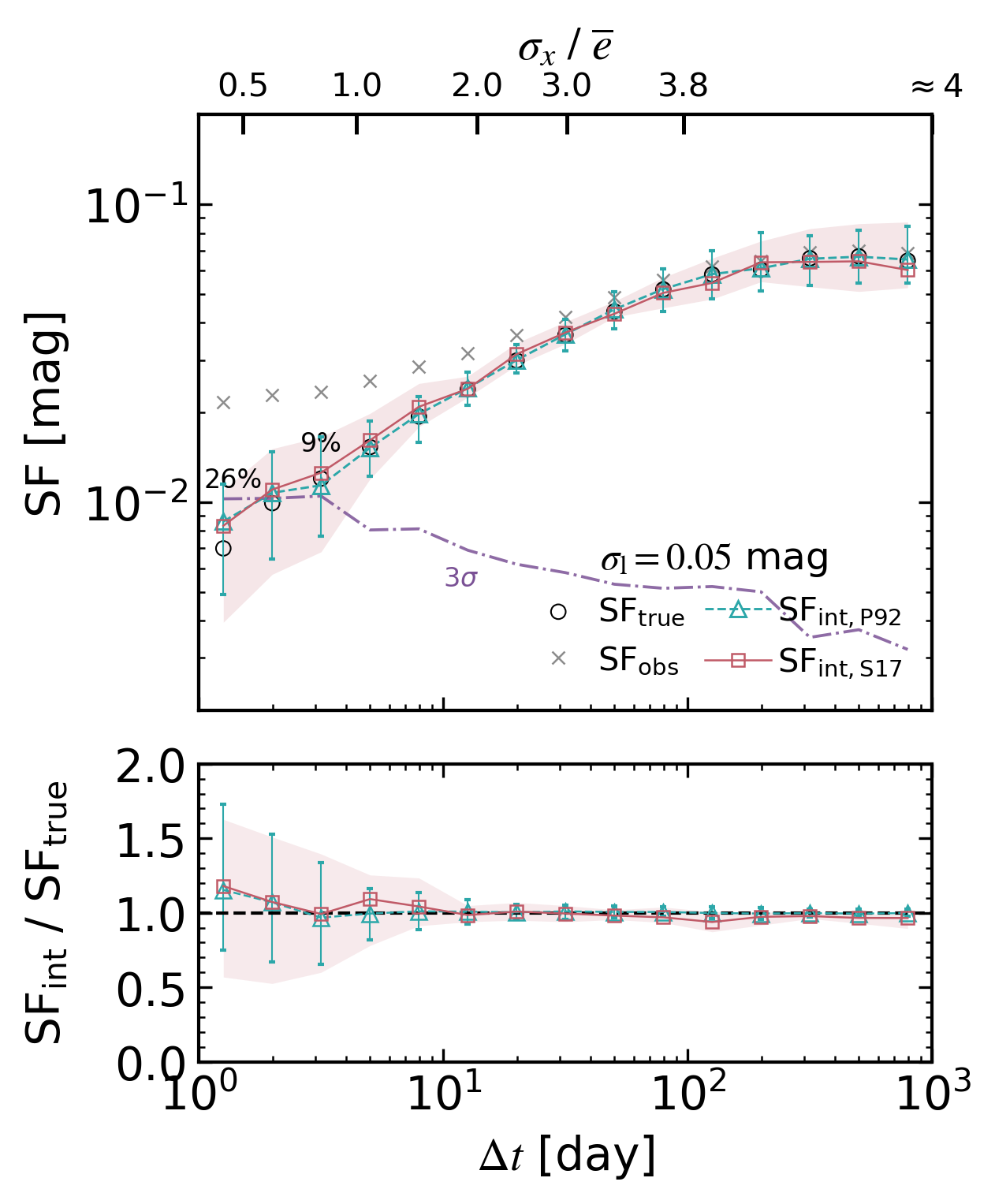} 
    \includegraphics[width=0.32\linewidth]{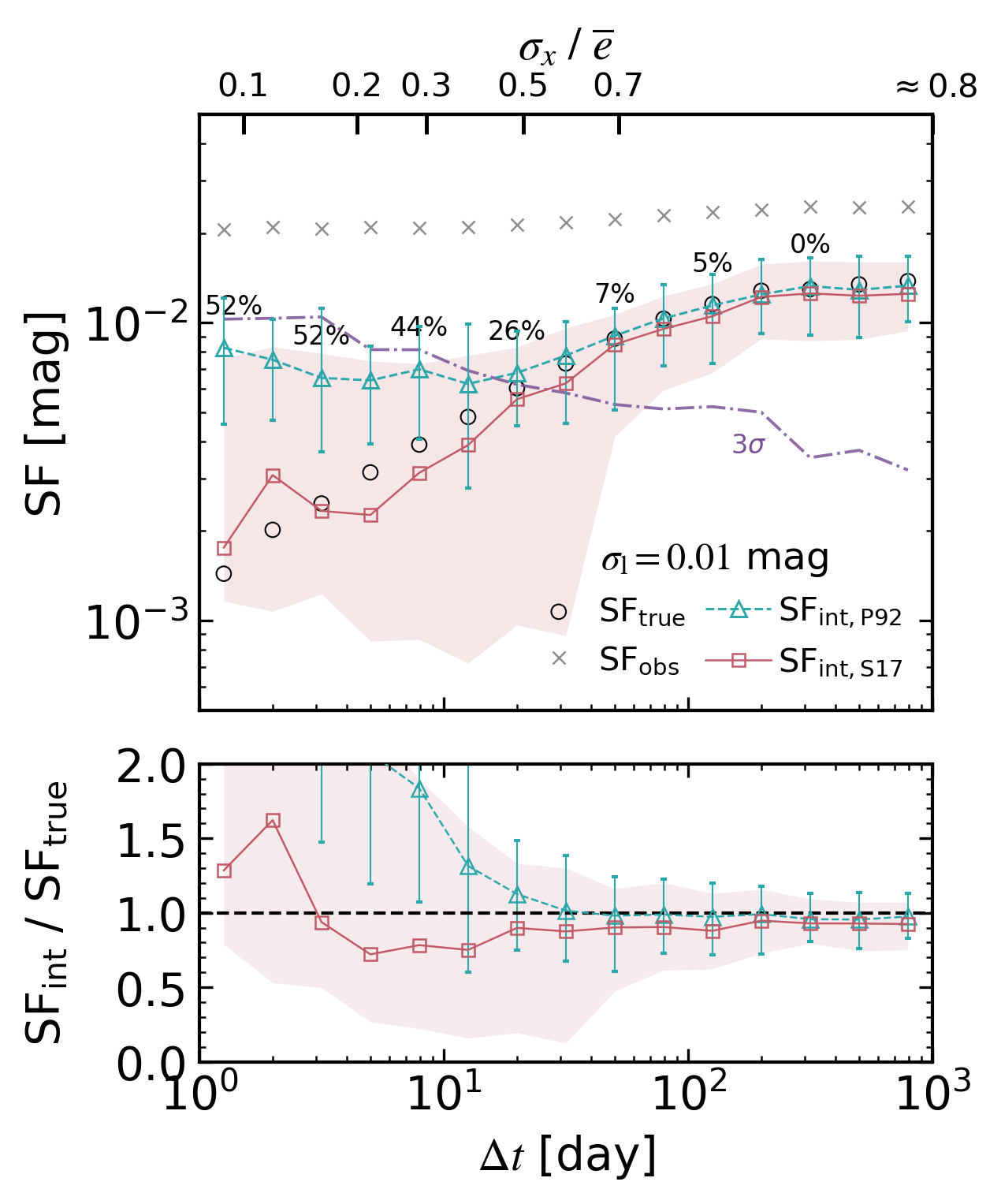} 
    \includegraphics[width=0.32\linewidth]{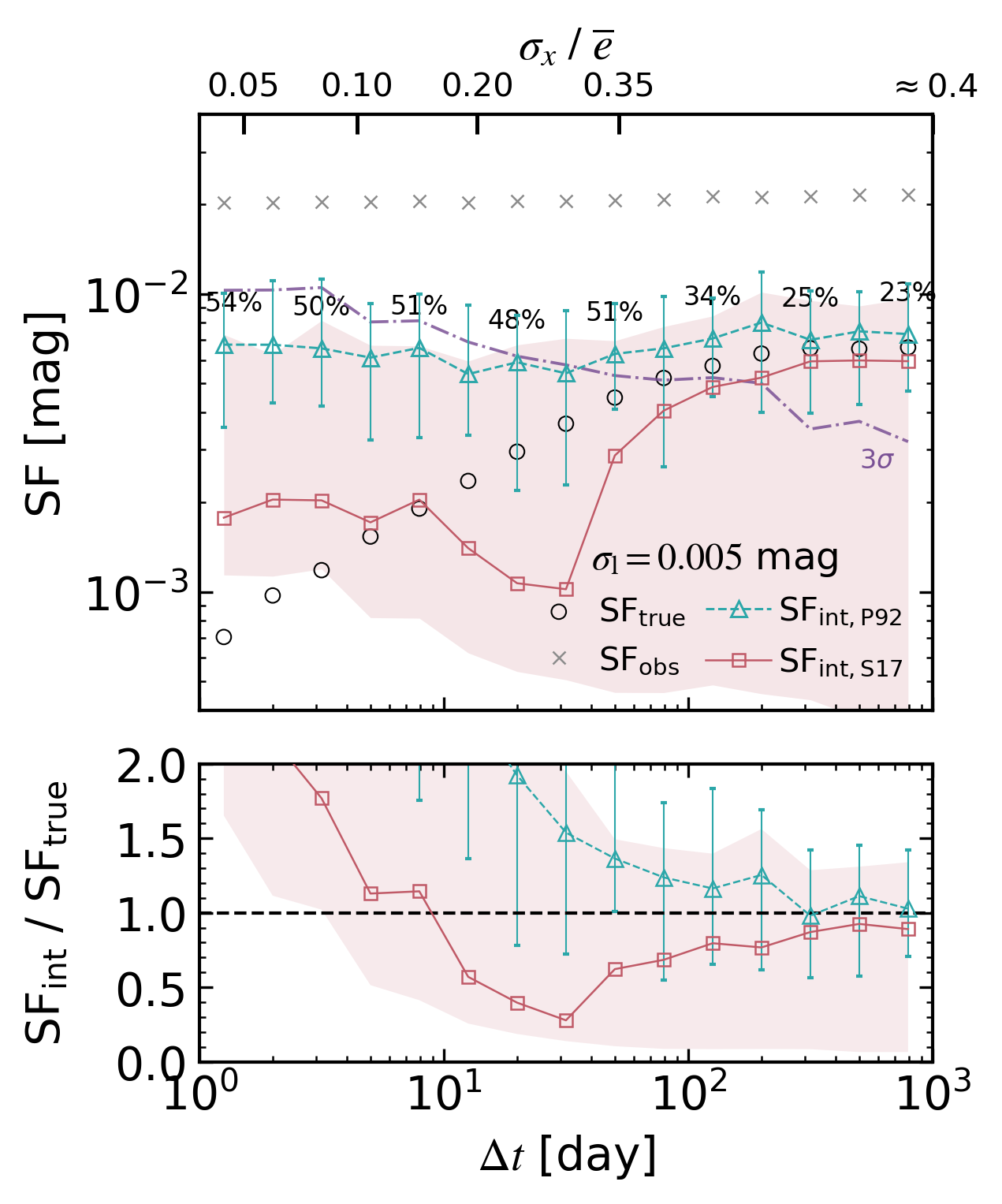}
    \caption{Similar to Figure~\ref{fig:p92_vs_s17}, but for light curves sampled using the Stripe 82 cadence illustrated in Figure~\ref{fig:sampling-lc}. The P92 and S17 SFs are calculated separately for each of 100 realizations; symbols show the median estimates, while error bars and shaded regions show the 16th--84th percentile ranges across realizations.
    The purple dash-dotted curve marks the idealized 3$\sigma$ detection threshold (Equation~\ref{eq:s17_thresh}), calculated from the number of pairs and photometric errors in each timescale bin.
    }
    \label{fig:sf_sampling}
\end{figure*}

\begin{figure}
    \centering
    \includegraphics[width=0.45\textwidth]{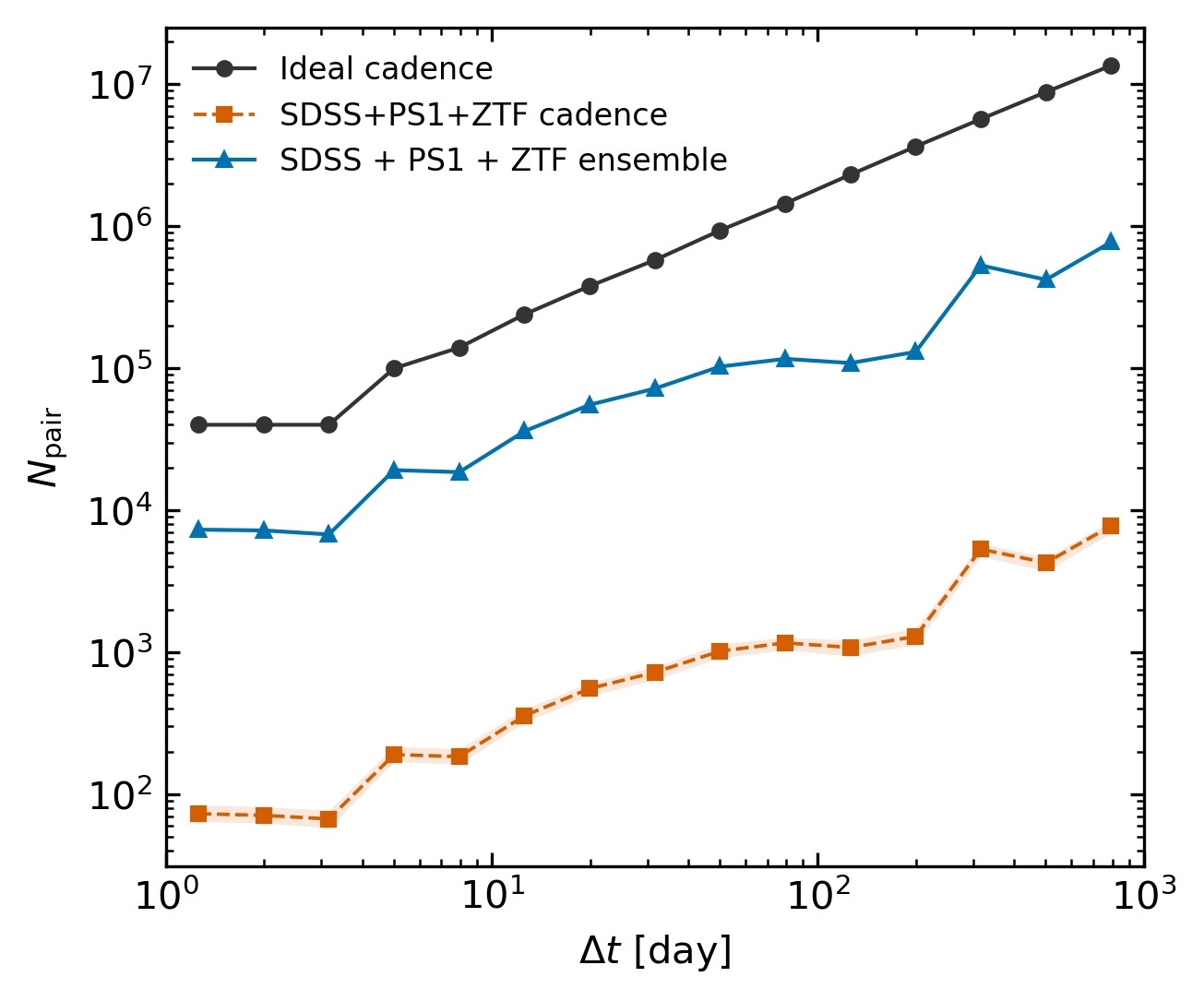}
    \caption{The dependence of $N_{\rm pair}$ on time interval for ideal and survey-sampled light curves. 
    Black circles show the median number of pairs across 100 ideally sampled light curves. Orange squares show the median number of pairs per light curve after applying the SDSS, PS1, and ZTF sampling. Blue triangles show the total number of pairs after combining the pair counts from all 100 sampled light curves in each time bin.}
    \label{fig:npair}
\end{figure}

\begin{figure*}
     \centering
    \includegraphics[width=0.32\linewidth]{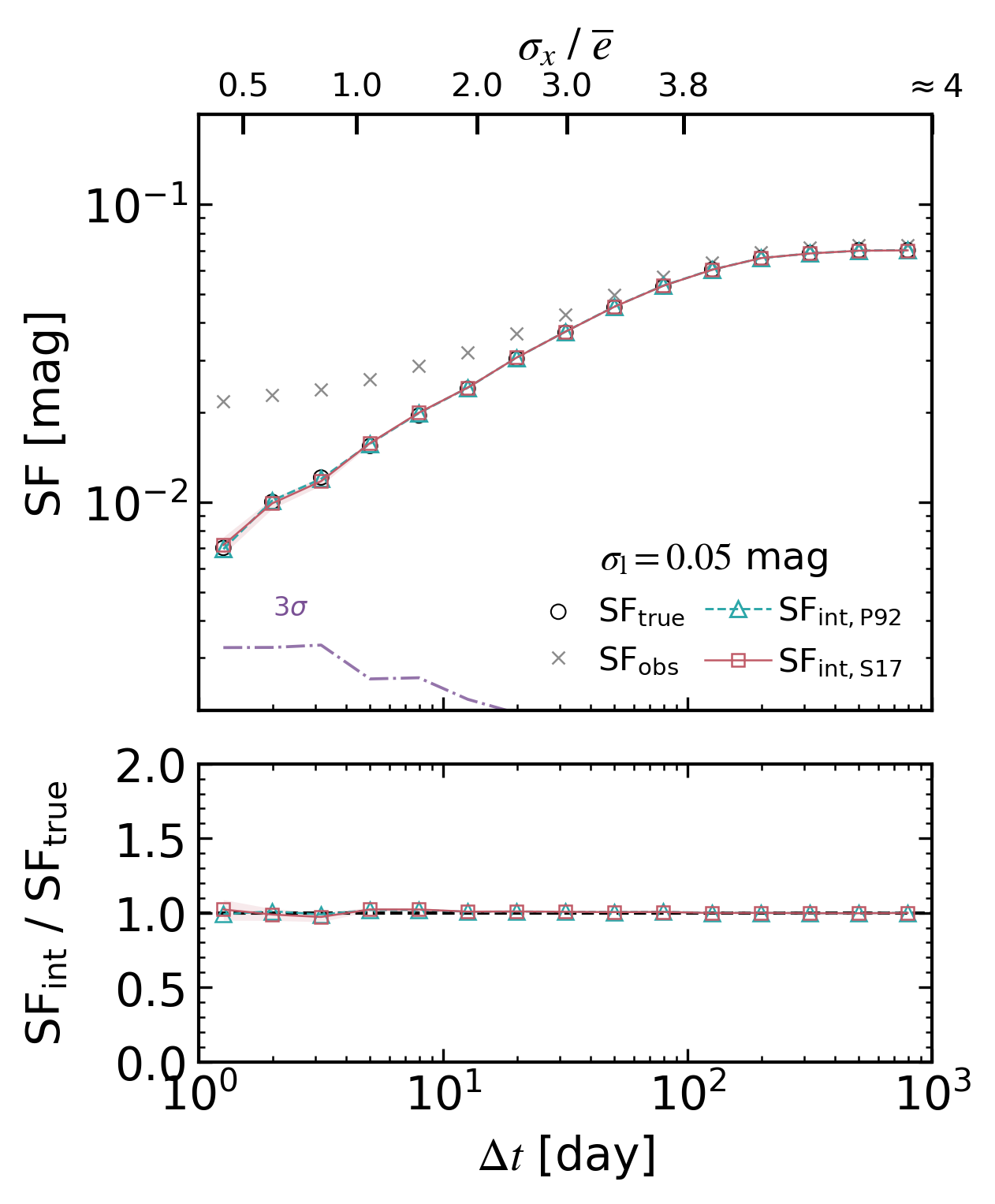}
    \includegraphics[width=0.32\linewidth]{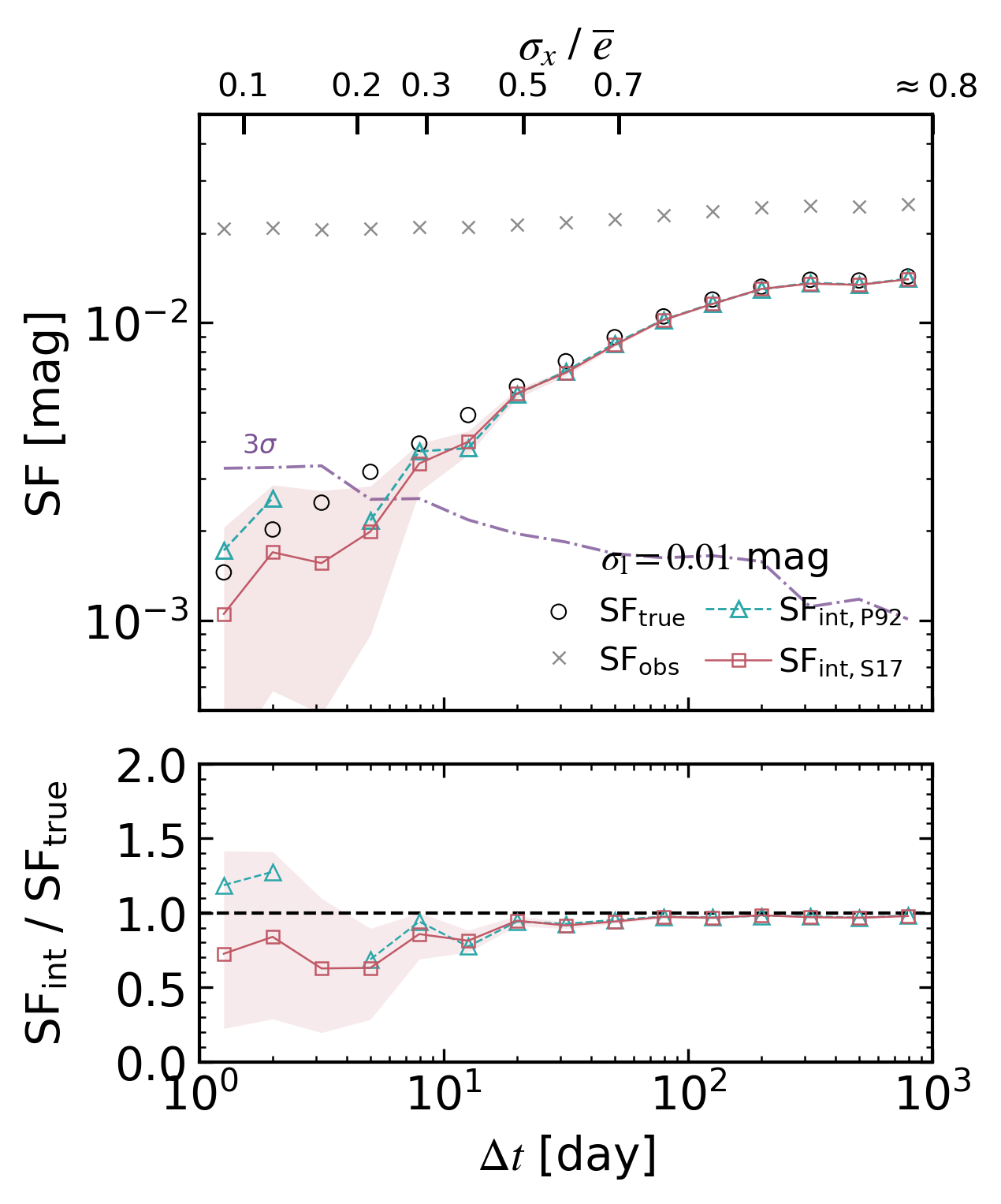}
    \includegraphics[width=0.32\linewidth]{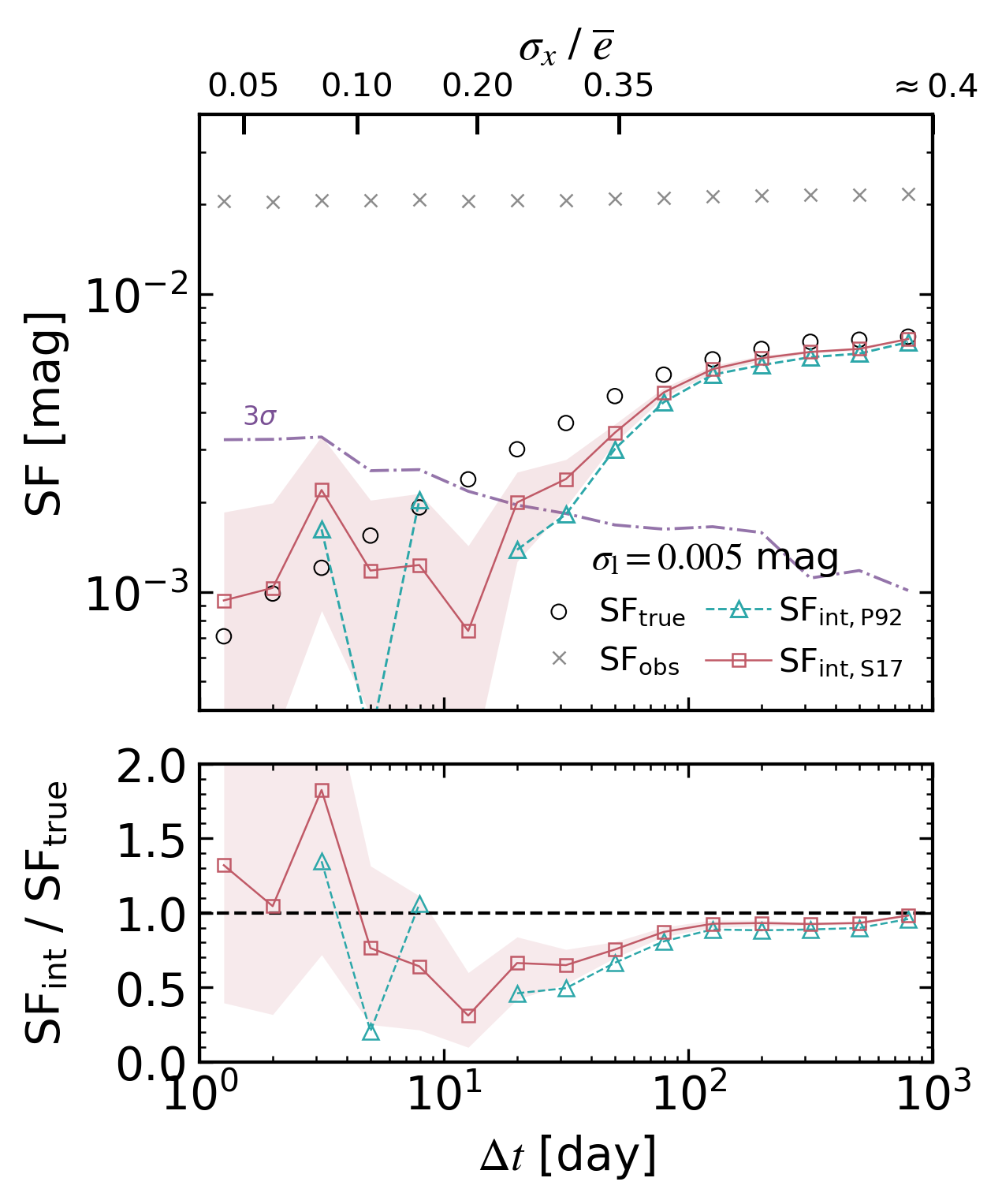   }
    \caption{Similar to Figure~\ref{fig:p92_vs_s17}, but for the ensemble SF calculated from 100 light curves sampled using the Stripe 82 cadence illustrated in Figure~\ref{fig:sampling-lc}. The magnitude-difference pairs from the 100 sources are combined in each timescale bin to obtain a single ensemble SF. 
    In the middle and right panels, the noise-subtracted ensemble ${\rm SF}_{\rm int,P92}^{2}$ is negative in some short timescale bins, so the corresponding P92 SF amplitudes cannot be shown.
    At high signal-to-noise ratios, the estimates generally recover the input SF; at low signal-to-noise ratios, departures from the true SF remain. 
    The purple dash-dotted curve marks the idealized 3$\sigma$ detection threshold (Equation~\ref{eq:s17_thresh}), calculated from the number of pairs and photometric errors in each timescale bin.}
    \label{fig:sf_sampling_ensemble}
\end{figure*}

\subsubsection{SF Recovery for Individual Light Curves}

Figure~\ref{fig:sf_sampling} shows the P92 and S17 estimates for the 100 sampled light curves.
Compared with the results for the ideally sampled light curves in Figure~\ref{fig:p92_vs_s17}, the recovery of intrinsic variability is less reliable after applying the Stripe 82 cadence.

For P92, the median estimates at $\sigma_{\rm l}=0.05$ mag (left panel) remain close to the input SF, although the 16th--84th percentile ranges are broader than in the ideal-sampling case.
At $\sigma_{\rm l}=0.01$ and 0.005~mag (middle and right panels), the estimates show a clear floor over much of the timescale range.
For S17, the median estimates at $\sigma_{\rm l}=0.05$ mag remain close to the input SF, but the percentile ranges are broader than in the ideal-sampling case. At $\sigma_{\rm l}=0.01$ mag, the median departs slightly from the input SF at short timescales and approaches it toward longer timescales, while the percentile ranges remain broad. At $\sigma_{\rm l}=0.005$ mag, the estimates vary substantially across realizations over a broad range of timescales, and their median departs from the input SF.

Figure~\ref{fig:npair} compares the median $N_{\rm pair}$ per light curve for the ideal (circles) and survey-sampled (squares) cases, helping explain the differences between Figures~\ref{fig:p92_vs_s17} and \ref{fig:sf_sampling}. In the first three timescale bins, the sampled light curves contain a median of $\simeq70$ pairs per bin, compared with $\simeq4\times10^4$ for the ideal light curves, a reduction by a factor of $\simeq570$. According to Equations~\eqref{eq:sffloor} and \eqref{eq:s17_thresh}, these reduced pair counts help explain the P92 floor and the larger scatter in the S17 estimates, respectively.

To assess how well weak variability can be recovered under survey sampling, we show the nominal $3\sigma$ detection threshold as purple dash-dotted curves in Figure~\ref{fig:sf_sampling}. We calculate this threshold from Equation~\eqref{eq:s17_thresh} with $n=3$, using the pair count and photometric errors in each timescale bin. At $\sigma_{\rm l}=0.05$~mag (left panel), the true SF lies near the threshold at the shortest timescales and rises well above it at longer timescales, where both estimators recover it closely. At $\sigma_{\rm l}=0.01$~mag (middle panel), the true SF falls below the threshold at short timescales, where P92 shows a floor and the S17 estimates have broad scatter. At $\sigma_{\rm l}=0.005$~mag (right panel), the true SF remains near or below the threshold over a wider timescale range, and recovery is correspondingly less reliable. 

It should be noted that the 3$\sigma$ threshold calculated from Equation~\eqref{eq:s17_thresh} serves as a reference and does not by itself guarantee reliable SF recovery. 
This is because $N_{\rm pair}$ estimated from the real observations are not completely independent as required by Equation~\eqref{eq:s17_thresh}. 
We therefore recommend assessing SF recovery with simulations matched to the cadence and photometric errors in real observations.

\subsubsection{SF Recovery from an Ensemble of Light Curves}

How can we improve the SF recovery? Equation~\eqref{eq:s17_thresh} suggests either improving photometric precision or increasing the number of independent pairs. For a given survey, however, photometric precision is limited by observing conditions and calibration systematics. Combining light curves therefore offers the only way to improve the SF recovery. 
Figure~\ref{fig:sf_sampling_ensemble} shows the ensemble SF, calculated by combining the magnitude-difference pairs from all 100 sampled light curves in each timescale bin. When $\sigma_{\rm l}>\bar{e}$ ($\sigma_{\rm l}=0.05$ mag; left panel), both P92 and S17 remain close to the input SF. At $\sigma_{\rm l}=0.01$ mag (middle panel), both estimates show a slight departure from the input SF at short timescales. The departure is larger at $\sigma_{\rm l}=0.005$ mag (right panel), particularly at short timescales. Thus, combining light curves improves the recovery, although departures from the input SF remain when the signal is weak.

For the Stripe 82 cadence, combining 100 light curves with similar physical properties indeed increases the pair count and lowers the idealized $3\sigma$ threshold, shown by the purple dash-dotted curves in Figure~\ref{fig:sf_sampling_ensemble}. However, the ensemble SF remains slightly biased when the intrinsic variability is weak. 
According to Figure~\ref{fig:npair}, in the shortest-timescale bin, the 100 sampled light curves (triangles) contribute about $7.3\times10^3$ pairs to the ensemble SF, compared with about $4\times10^4$ pairs for one ideally sampled (circles) light curve. Under the idealized assumption that the pair count increases in proportion to the number of similar sources, $\simeq550$ quasars with similar physical variation properties would be needed to reach the ideal-light-curve pair count in this bin. 
This estimate suggests that larger samples plus denser observing cadence are needed to improve recovery of weak variability.

\subsection{Beyond the DRW process}\label{sec:beyond_drw}

\begin{figure*}[!t]
    \centering
    \includegraphics[width=0.4\linewidth]{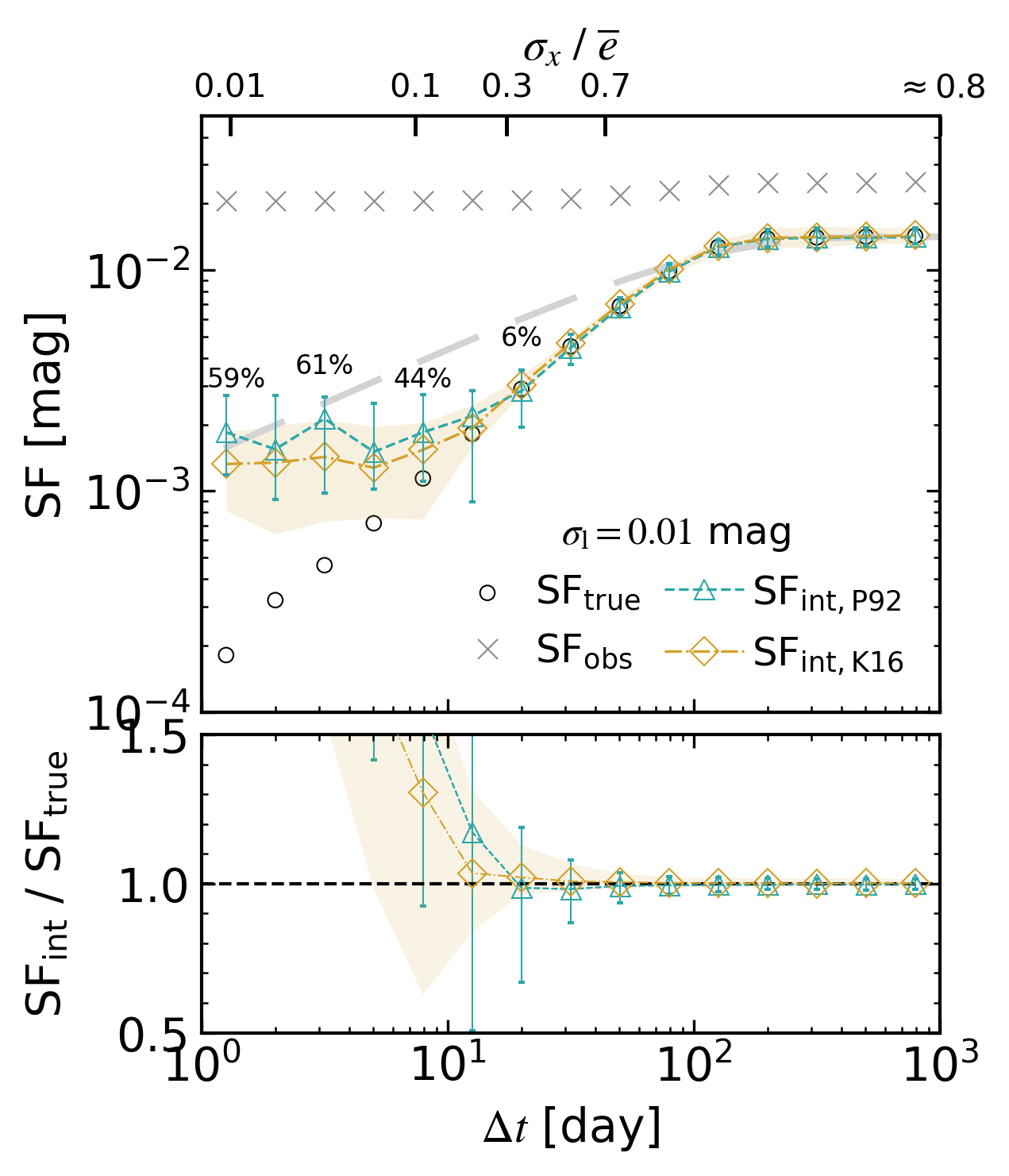}
    \includegraphics[width=0.4\linewidth]{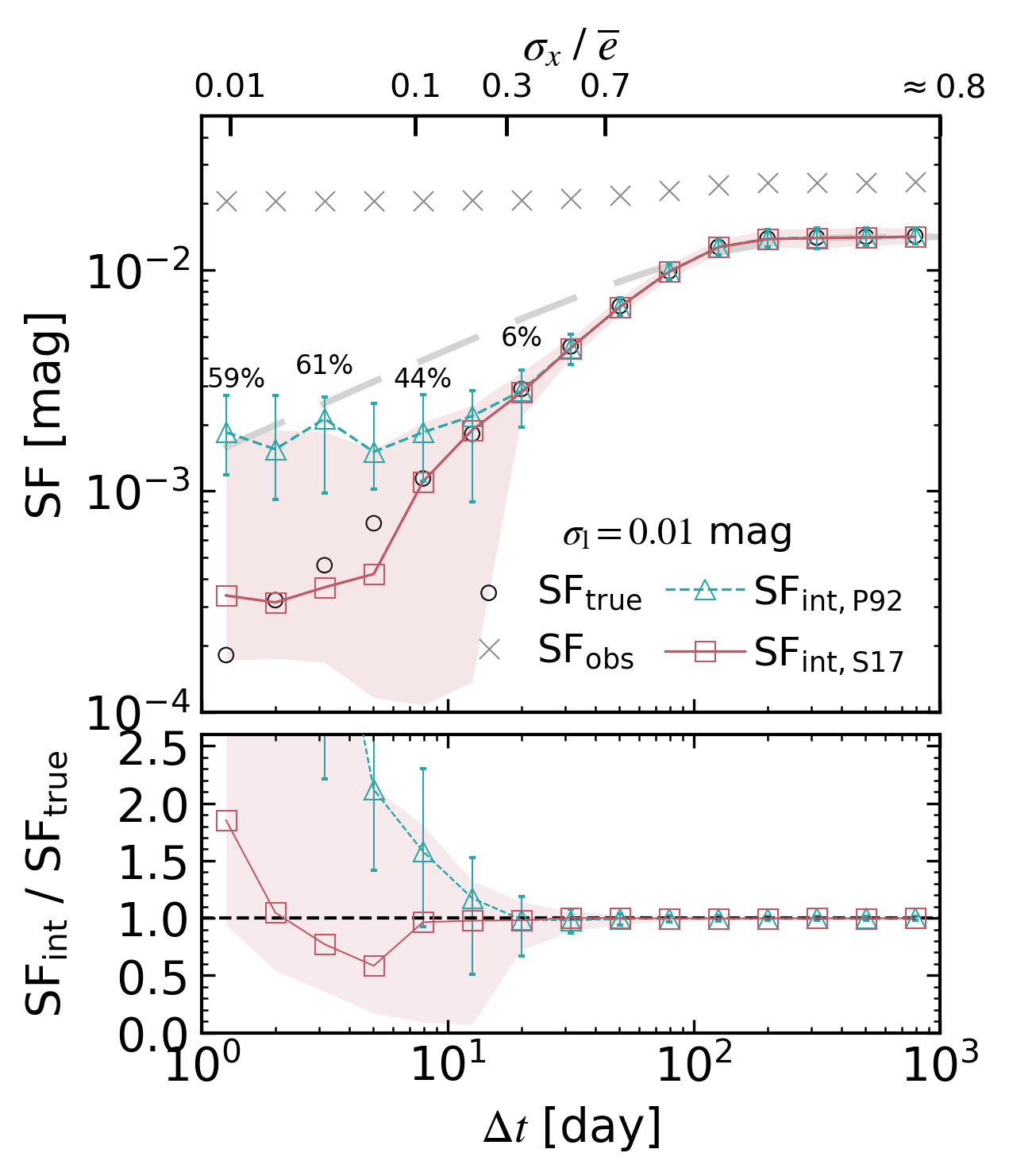}
    \caption{Recovery of the intrinsic SF for an non-DRW process with $\beta=2$ and $\sigma_{\rm l}=0.01$~mag. 
    Analogous to panels of Figures~\ref{fig:p92_vs_k16} and~\ref{fig:p92_vs_s17}, the left column compares the P92 (triangles) and K16 (diamonds) estimators, while the right column the P92 (triangles) and S17 (squares) estimators. For reference, the grey long-dashed line marks the DRW case with $\beta=1$ and $\sigma_{\rm l}=0.01$~mag. 
    }
    \label{fig:non-drw}
\end{figure*}

\begin{figure*}[!t]
    \centering
    \includegraphics[width=0.4\linewidth]{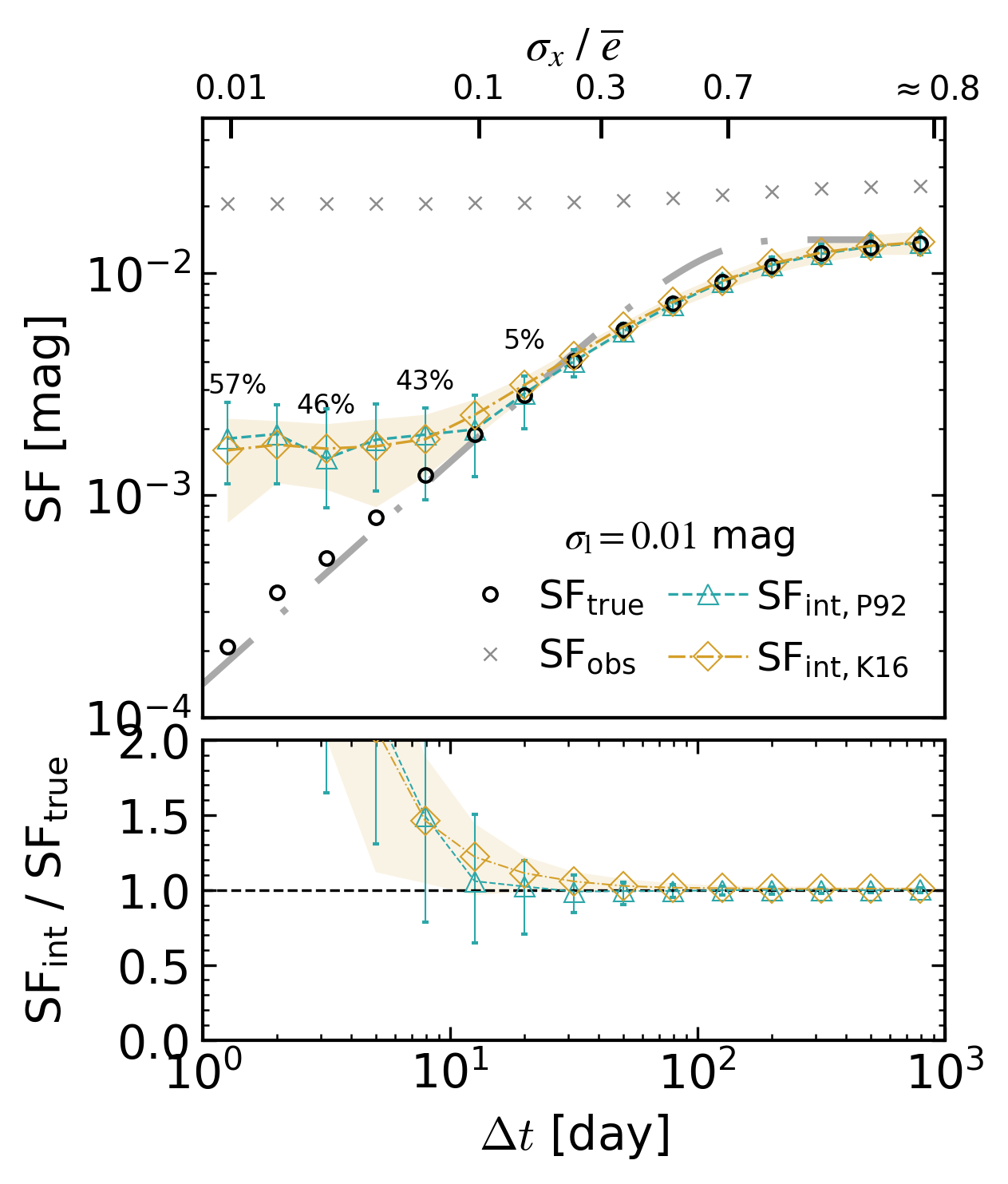}
    \includegraphics[width=0.4\linewidth]{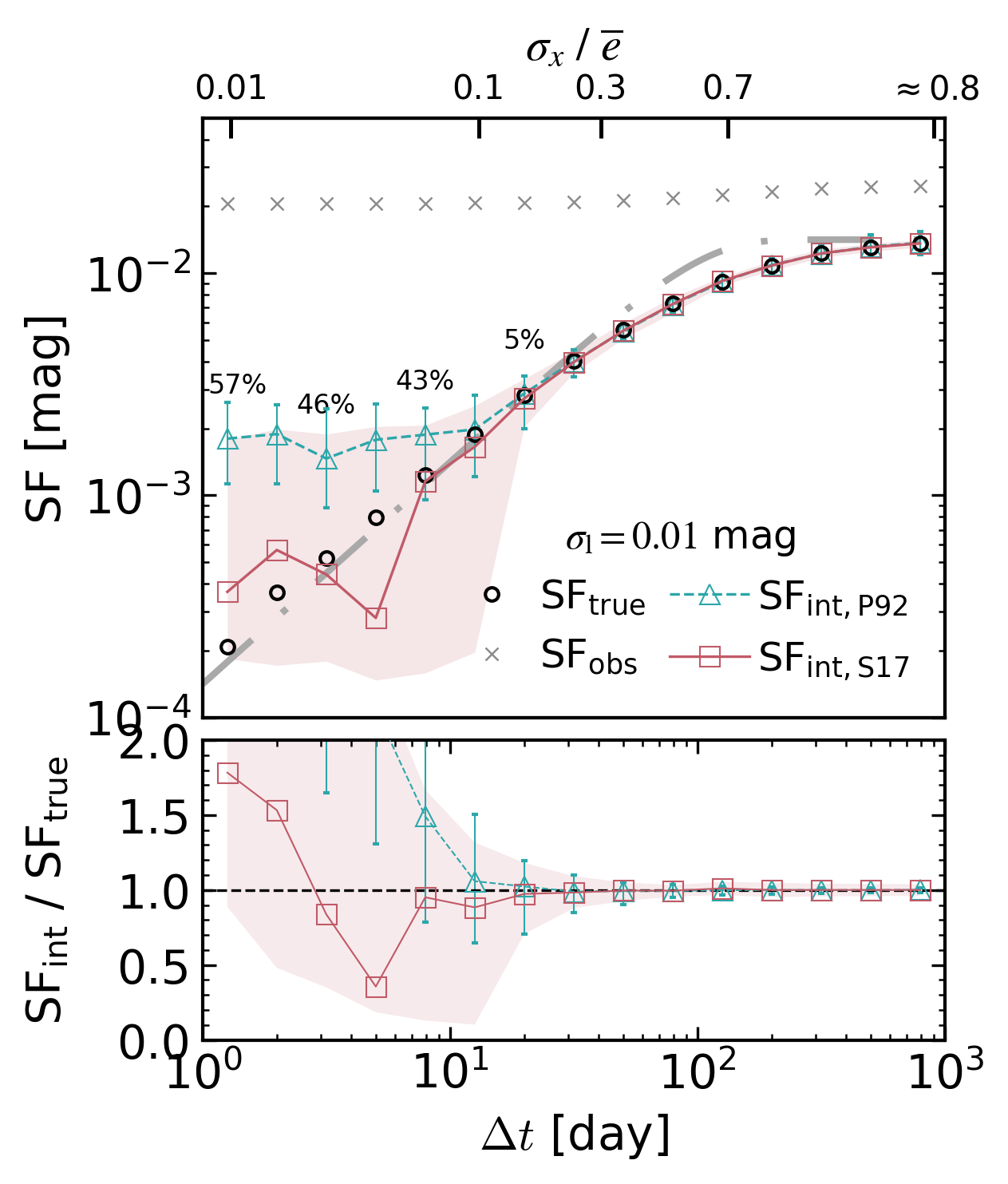}
    \caption{Similar to Figure~\ref{fig:non-drw}, but for a CARMA$(4,2)$ process with $\sigma_{\rm l}=0.01$~mag. The left column compares the P92 (triangles) and K16 (diamonds) estimators, and the right column compares the P92 (triangles) and S17 (squares). The black circles show the median SF calculated from the ideal CARMA$(4,2)$ light curves without photometric errors.
    In both panels, the dash-dotted curves show the SFs corresponding to the power-exponential covariance models in Equation~\eqref{eq:cov} with $\beta=2$.}
    \label{fig:carma42_p92_s17}
\end{figure*}

Herein, we have compared the P92, K16, and S17 estimators for a DRW ($\beta=1$) intrinsic process, finding S17 to be the best among the three (Figures~\ref{fig:p92_vs_k16} and~\ref{fig:p92_vs_s17}). Optical AGN variability, however, is found to depart from a DRW both at short timescales \citep[e.g.,][]{Stone2023,Mushotzky2011ApJ...743L..12M,Simm2016A&A...585A.129S,Su2024a,Petrecca2025,Kasliwal2015,Smith2018ApJ...857..141S,Arevalo2024} and at long timescales \citep[e.g.,][]{Guo2017ApJ...847..132G,Petrecca2025,Arevalo2024}, pointing to a stochastic process more complex than the DRW. 
At long timescales the variability amplitude is large and generally well above the photometric error, so such departures are easily measured; at short timescales the amplitude is small and comparable to the error, which is where noise treatment matters most. We therefore focus on the short-timescale regime, and check whether S17 remains the best among the three estimators.

To test the non-DRW case, we first consider a power-exponential model with $\beta=2$ as an example (Equation~\ref{eq:cov}), generate $100$ realizations (same cadences and parameters as in Section~\ref{subsec:parameter_selection}, except $\beta$), and apply all three estimators. 
As shown in Figure~\ref{fig:non-drw} for $\sigma_{\rm l}=0.01$~mag, at long timescales ($\Delta t\gtrsim10$~d, or $\sigma_x/\bar e\gtrsim0.1$) all three recover the true SF accurately, with ${\rm SF}_{\rm int}/{\rm SF}_{\rm true} \approx 1$.
At short timescales, however, recovery becomes harder than in the DRW case because the steeper $\beta=2$ SF is weaker there, yielding a lower signal-to-noise ratio at fixed photometric error. 
The estimators behave similarly to the DRW case, but with notable differences: 
P92 reaches a floor that affects a broader range of timescales; K16 suffers from stronger degeneracy; and S17 remains nearly unbiased$-$its median deviates only at the shortest timescales, while its uncertainties broaden, reflecting a loss of precision rather than accuracy, consistent with its near-optimal behaviour (Section~\ref{sec:s17_optimal}).

We next extend simulations beyond the covariance formalism of Equation~\eqref{eq:cov} by generating light curves with the CARMA process. In a CARMA$(p,q)$ model, $p$ and $q$ denote the autoregressive and moving-average orders, respectively, with $p>q$. 
The DRW is a CARMA$(1,0)$ process, the continuous-time analogue of an AR(1) model, and has a single characteristic decorrelation timescale. Higher-order CARMA processes can produce a wider range of PSD shapes \citep{Kelly2014}.
Fitting CARMA models to quasar light curves, \citet{Stone2023} found that the preferred orders cluster around $p\approx4$ and $q\approx2$.
They also showed that a forced CARMA$(2,1)$ model can produce PSDs similar to those of the best-fit higher-order CARMA models.
Motivated by \citet{Stone2023}, we test the SF estimators based on the CARMA$(4,2)$ process.

We generate 100 ideal light curves using the state-space method of \citet{Kelly2014} and adopting the CARMA parameters of the source with DBID$=$2552293 fit by \citet[][cf. their Table~1]{Stone2023}. We normalize these light curves to have $\bar m = 17$~mag and a stationary standard deviation of $\sigma_{\rm l} =0.01$~mag, adding photometric errors as in Section~\ref{sec:sim_lc}.

Figure~\ref{fig:carma42_p92_s17} shows the SF recovery for 100 CARMA$(4,2)$ realizations. At long timescales, all three estimators recover the true SF closely. At short timescales, however, recovery becomes more difficult as the intrinsic signal weakens relative to the photometric error. P92 reaches a noise floor, while K16 also shows substantial departures from the true SF at short timescales. S17 remains generally unbiased, with its median departing from the true SF only at the shortest timescales, while its uncertainties broaden as expected from Equation~\ref{eq:s17_thresh}. This behavior is consistent with its near-optimal performance discussed in Section~\ref{sec:s17_optimal}.

In Figure~\ref{fig:carma42_p92_s17}, the median SF calculated from the ideal CARMA$(4,2)$ light curves (black circles) is close to that generated using the covariance function in Equation~\ref{eq:cov} with $\beta=2$ (dash-dotted curves) at short timescales, although they differ at long timescales.
The CARMA$(4,2)$ PSD asymptotically follows $P(f)\propto f^{-4}$ at high frequencies, as follows from Equation~(3) of \citet{Kelly2014}, implying ${\rm SF}\propto\Delta t$ in the short-timescale limit.\footnote{The usual relation $\alpha=2\gamma+1$, for $P(f)\propto f^{-\alpha}$ and ${\rm SF}\propto\Delta t^\gamma$, applies only for $1<\alpha<3$ \citep{Sartori2019}. For a one-sided PSD, ${\rm SF}^2(\Delta t)=2\int_0^\infty P(f)[1-\cos(2\pi f\Delta t)]\,df$. For the stationary CARMA$(4,2)$ process, the high-frequency behavior $P(f)\propto f^{-4}$ ensures that $\int_0^\infty f^2P(f)\,df$ is finite. Expanding the cosine in the short-timescale limit therefore gives ${\rm SF}^2(\Delta t)\simeq4\pi^2\Delta t^2\int_0^\infty f^2P(f)\,df$, implying ${\rm SF}\propto\Delta t$.
Their similar recovery behavior may therefore reflect the comparable intrinsic signals relative to the photometric errors, consistent with the dependence found in the previous simulations.}

Our assessment therefore extends beyond the DRW case. Given accurate input photometric errors, S17 provides the most faithful recovery of the intrinsic SF and the most reliable uncertainty estimates, regardless of whether the underlying process is DRW or non-DRW.

\subsection{Beyond well-calibrated photometric errors}\label{sec:photo_errors}

\begin{figure*}[!t]
    \centering
    \includegraphics[width=0.4\linewidth]{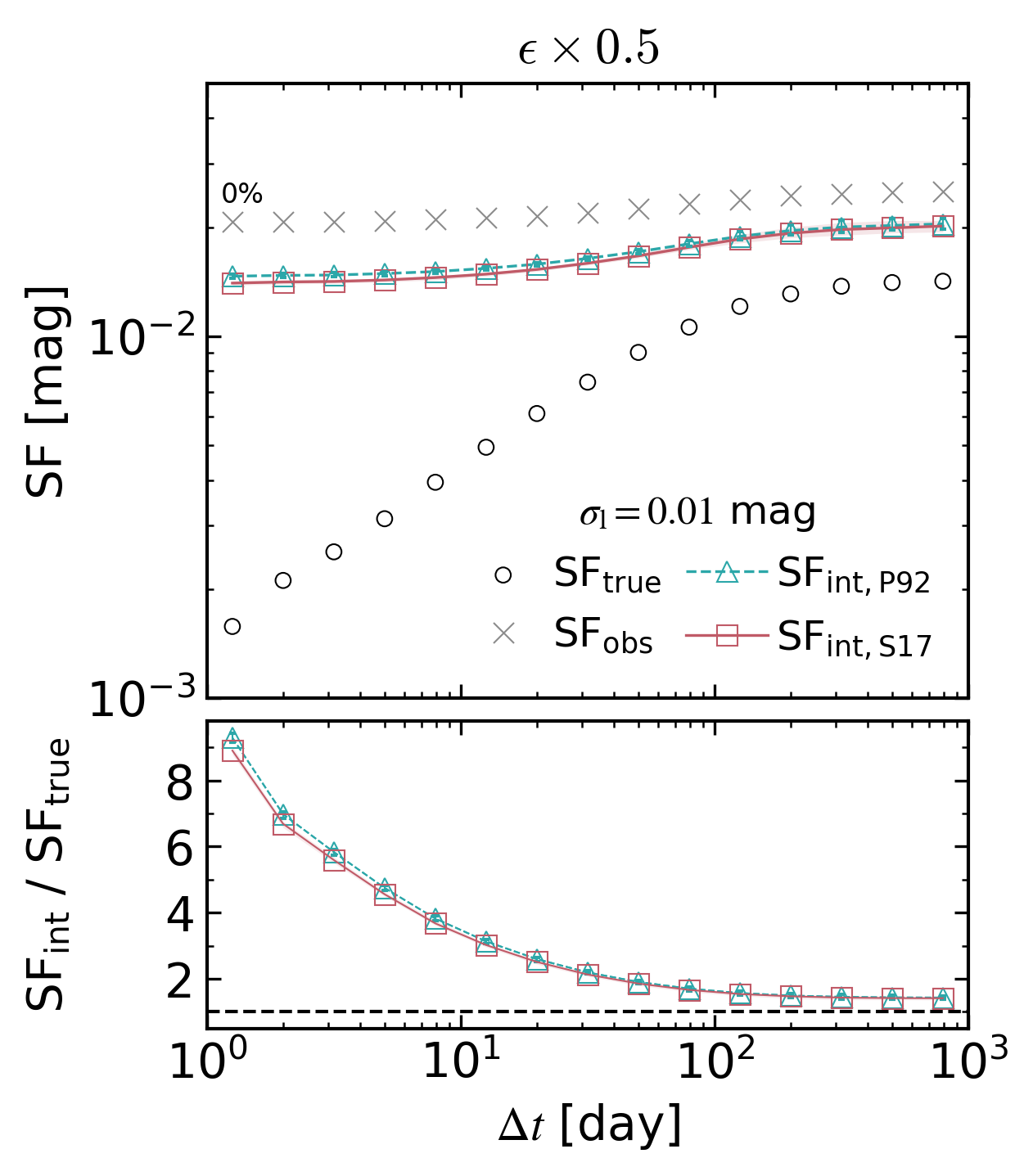}
    \includegraphics[width=0.4\linewidth]{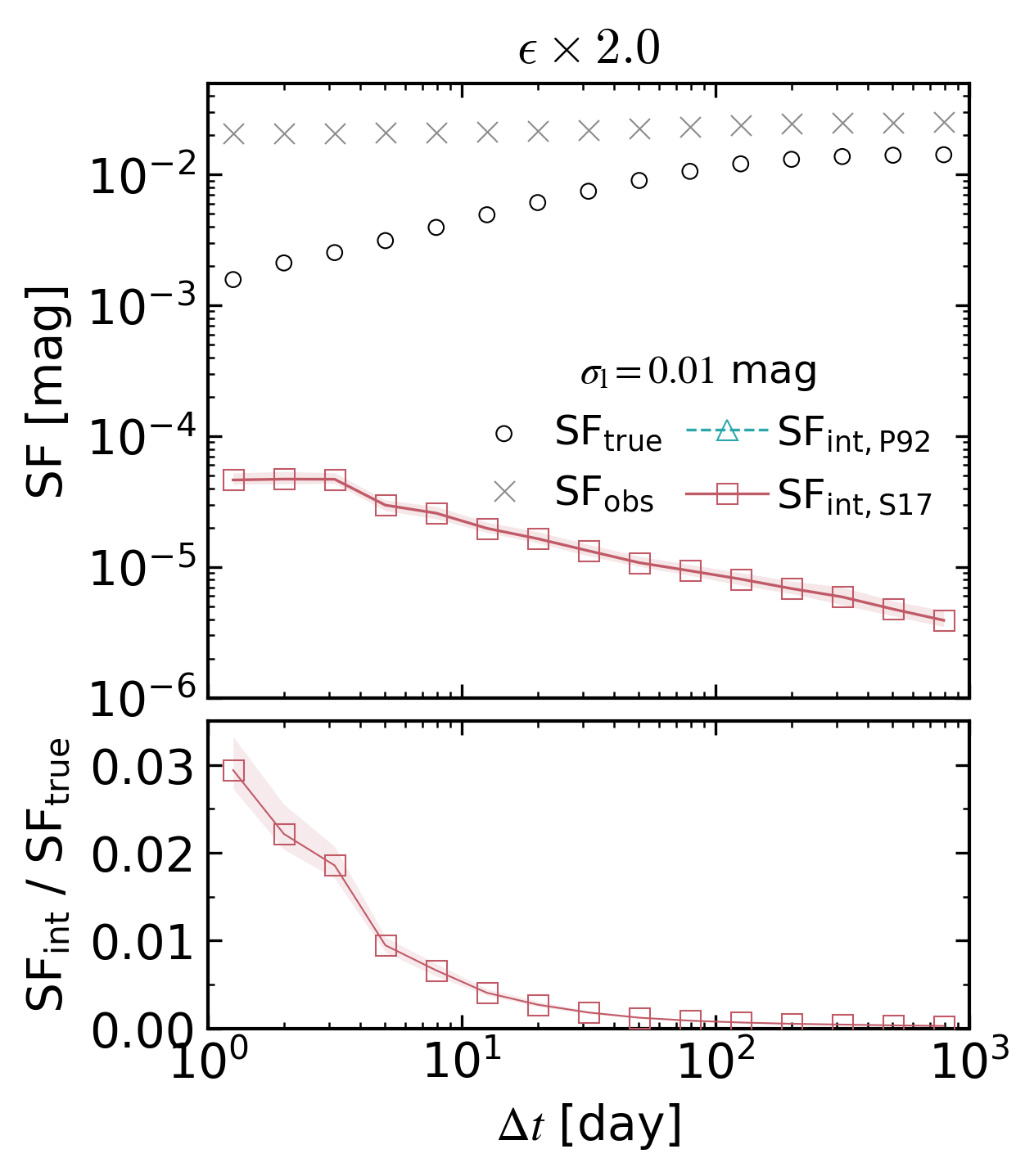}
    \caption{Effect of mis-estimating photometric error on the recovery of the intrinsic SF given $\sigma_{\rm l}=0.01$~mag. The photometric errors are underestimated (overestimated) by a factor of two in the left (right) column.
    When errors are underestimated, the noise is incompletely removed and both SF estimators overestimate the intrinsic SF, biasing $\mathrm{SF}_{\mathrm{int}}/\mathrm{SF}_{\mathrm{true}}$ high. 
    When errors are overestimated, the noise is over-subtracted, driving the S17 estimate toward zero so that $\mathrm{SF}_{\mathrm{int,S17}}/\mathrm{SF}_{\mathrm{true}}$ falls far below unity. For P92, noise over-subtraction yields negative ${\rm SF}_{\rm int,P92}^{2}$ values, so the corresponding SF amplitudes cannot be shown.
    Therefore, having accurate photometric errors is necessary for recovering the intrinsic SF with the S17 estimator.
    }
    \label{fig:uncertainties}
\end{figure*}

So far, we have demonstrated the performances of three SF estimators and identified S17 as the best performer. It should be noted, however, that the accuracy of S17 (and that of P92) depends critically on the accuracy of the measured photometric errors. In reality, photometric errors are frequently not well characterized \citep[e.g.,][]{Wyrzykowski2009,Skowron2016,Vaughan2016,Suberlak2017,Goto2021,Zhang+2025}. This is often the case for bright sources, where errors are under-estimated because only statistical uncertainties are reported and systematic contributions are not properly considered. 
For example, as illustrated in Figure~\ref{fig:e_sigma_e_mag}, directly using the photometric errors reported by \citet[][diamonds]{MacLeod2010ApJ...721.1014M} for AGNs brighter than $m \simeq 19$~mag, the corresponding mean photometric errors are systematically smaller than those inferred from the non-variable stars constructed by \citet[][circles]{Ivezic2007}, indicating that the photometric errors reported for bright AGNs are under-estimated.
Contrarily, the mean photometric errors of AGNs fainter than $m \simeq 19$~mag are slightly larger than the those inferred from the non-variable stars, indicating that the photometric errors reported for faint AGNs are over-estimated (Figure~\ref{fig:e_sigma_e_mag}).
Similar under-estimation (over-estimation) of photometric errors for bright (faint) sources can be found in, e.g., \citet[][top panel of their Figure~A1]{Suberlak2017} and \citet[][top panel of their Figure~7]{Goto2021}.

Based on our 100 DRW realizations of $\sigma_{\rm l} = 0.01$~mag, we scale the photometric errors in all epochs by a constant factor $f$ to mimic under-/over-estimation, and evaluate the performance of P92 and S17.
Under-estimated errors ($f=0.5$, left panel of Figure~\ref{fig:uncertainties}) create a persistent positive bias that diminishes as the signal strengthens. Over-estimated errors ($f=2$, right panel of Figure~\ref{fig:uncertainties}), however, are far more severe: the negative bias can suppress the recovered variance entirely. For example, over-subtracting the noise can produce negative ${\rm SF}_{\rm int,P92}^{2}$ values in all timescale bins, in which case no real-valued P92 SF amplitude can be computed and shown in the right panel of Figure~\ref{fig:uncertainties}. The S17 estimates, by contrast, are driven toward zero, suggesting even S17 fails when errors are systematically over-estimated.

One might ask whether the bias could be removed simply by extending the S17 likelihood with $\sigma_i^2(\Delta t) = {\rm SF}_{\rm int,S17}^2(\Delta t) + (f\epsilon_i)^2$ and treating $f$ as a free parameter. This, however, leads to a degeneracy: within a single $\Delta t$ bin the data constrain only the sum ${\rm SF}_{\rm int,S17}^2 + (f\epsilon_i)^2$, so $f$ cannot be easily separated from ${\rm SF}_{\rm int,S17}$. 
This degeneracy could be broken by fitting a single global $f$ to all $\Delta t$ bins simultaneously since $f$ is timescale-independent while ${\rm SF}_{\rm int,S17}(\Delta t)$ is not.
A similar joint constraint on photometric errors and intrinsic variability has recently been demonstrated using Gaussian processes by \citet{Lefkir2025}, who fit a global error rescaling factor directly from the light curve.
A worthwhile next step would be to develop an upgraded SF estimator that simultaneously constrains the photometric errors and the intrinsic variability, and to evaluate its performance across different variability regimes. Such an investigation, however, lies outside the scope of this study.

Realistic photometric errors therefore require empirical recalibration using non-variable stars with similar magnitudes and in the same field \citep[e.g.,][]{Suberlak2017}. With accurate calibration, S17 recovers the intrinsic SF without significant bias; otherwise, both S17 and P92 are unreliable.

\begin{figure*}
    \centering
    \includegraphics[width=0.95\textwidth]{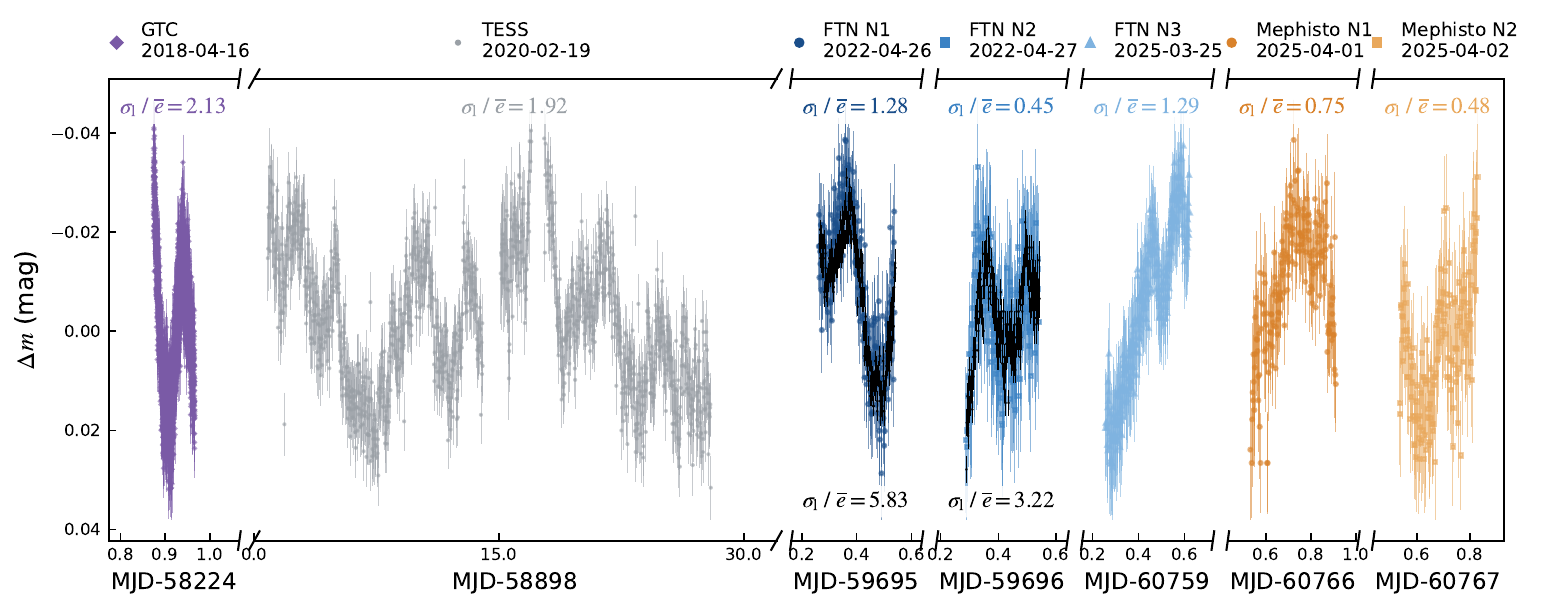}
    \caption{Median-magnitude-subtracted light curves of NGC~4395 ordered by observation date (GTC: \citealt{McHardy2023MNRAS.519.3366M}; TESS: \citealt{Burke2020ApJ...899..136B}; FTN: \citealt{Montano2022} and \citealt{Pan2026}; Mephisto: \citealt{Pan2026}).
    For FTN N1 and N2, measurements of \citet{Montano2022} and \citet{Pan2026} are in black and coloured symbols, respectively.
    The $\sigma_{\rm l}/\bar e$ values are indicated, where $\sigma_{\rm l}$ is the intrinsic variability amplitude estimated from the weighted excess variance of each dataset and $\bar e$ is the mean photometric error. The lower two values correspond to the measurements of \citet{Montano2022}.
    }
\label{fig:lc_ngc4395}
\end{figure*}

\subsection{Application to NGC 4395}\label{sec:ngc4395}

\begin{figure*}
    \centering
    \includegraphics[width=0.75\textwidth]{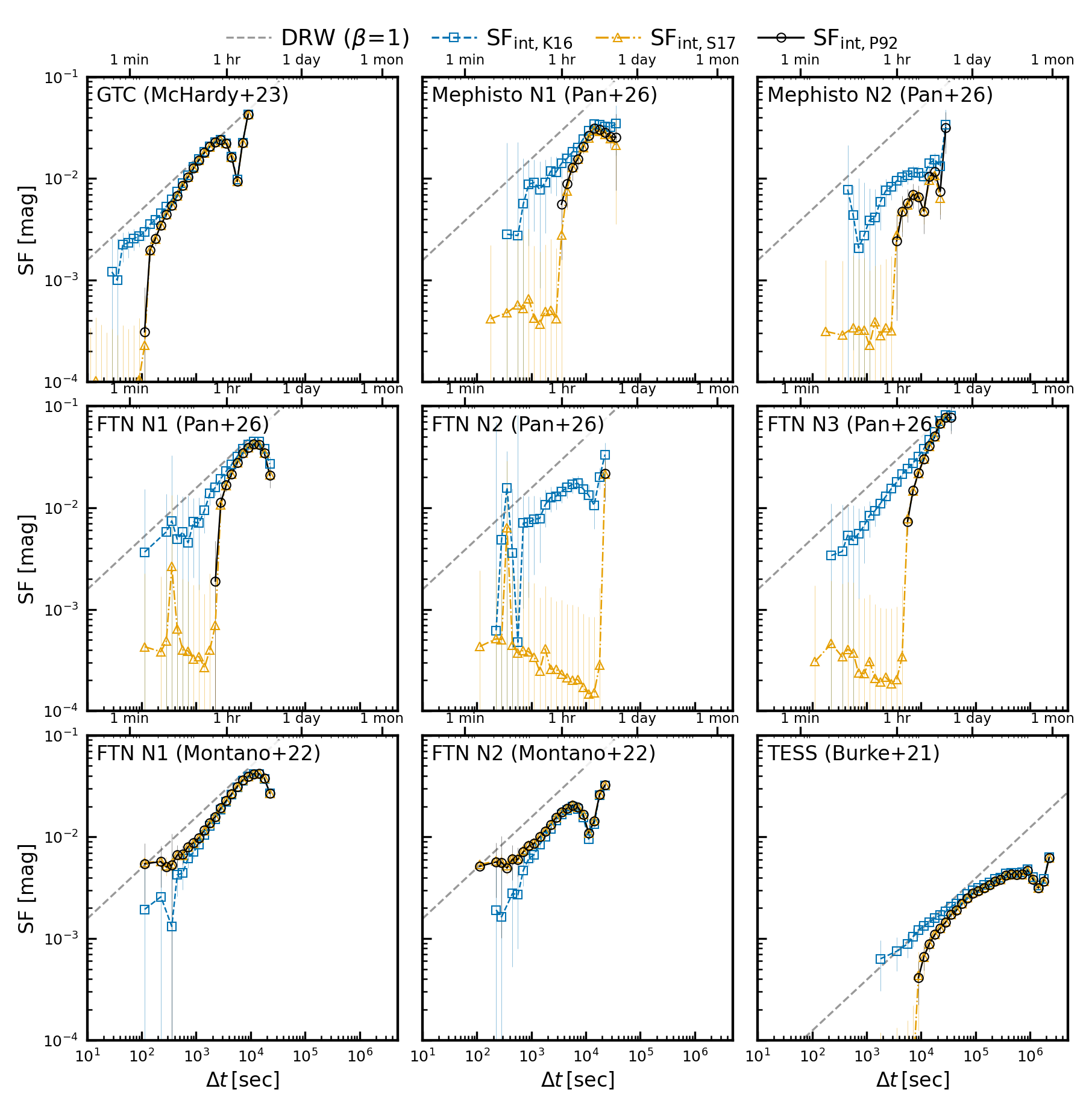}
    \caption{
    Intrinsic SFs of NGC 4395 recovered using the P92 (circles), K16 (squares), and S17 (triangles) estimators. 
    Error bars represent the $16$--$84$th percentile ranges, derived from the FR/RSS method for P92 and K16, and from the MCMC posterior for S17. 
    The grey dashed line is the DRW prediction ($\beta=1$).
    }
\label{fig:sf_ngc4395}
\end{figure*}

NGC~4395 provides a representative test of recovering weak variability from real data. Independent monitoring from multiple facilities spans timescales from seconds to months: high-cadence observations were obtained with the 10.4-m Gran Telescopio Canarias over one night \citep[GTC;][]{McHardy2023MNRAS.519.3366M}; TESS monitored the source for a month \citep{Burke2020ApJ...899..136B}; and ground-based monitoring used the Faulkes Telescope North \citep[FTN;][]{Montano2022,Pan2026} over three nights and the Multi-channel Photometric Survey Telescope \citep[Mephisto; ][]{Yuan2020} over two nights \citep{Pan2026}.
The FTN data were reduced independently using two methods: two nights with aperture photometry \citep{Montano2022}, and all three nights with image-subtraction photometry \citep{Pan2026}. The latter was also applied to Mephisto.

Together, these observations provide nine datasets (Figure~\ref{fig:lc_ngc4395}), each analysed separately, as differences in facilities and photometry prevent a simple combination into a single ensemble SF.
For each dataset, we estimate the variation amplitude over its observing baseline, denoted here by $\sigma_{\rm l}$, as the square root of the weighted excess variance and calculate its ratio to the mean photometric error, $\bar e$. The resulting $\sigma_{\rm l}/\bar e$ values are shown in Figure~\ref{fig:lc_ngc4395}.
We apply the three SF estimators to each dataset (Figure~\ref{fig:sf_ngc4395}).
Uncertainties for the recovered intrinsic SF are estimated via flux randomization/random subset selection \citep[FR/RSS;][]{Fausnaugh2016ApJ...821...56F} for P92 and K16, and from the MCMC posterior for S17.

At long timescales, P92 and S17 give consistent results. At short timescales, however, P92 often fails to provide useful information, whereas S17 does return estimates, albeit with very large uncertainties. Both estimators rely on accurate photometric errors; here, the reported errors are used as given, without recalibration against non‑variable stars, and are likely over‑ or under‑estimated across different datasets. Consequently, the P92 and S17 results cannot be interpreted reliably. Although many of their recovered SFs do show a short-timescale steepening relative to the DRW, this deviation must be treated with caution.

K16, by contrast, infers the noise level internally and is therefore unaffected by error mis-calibration. Following \citet{Su2024a}, we apply K16 to measurements at timescales shorter than the first local maximum of the observed SF. The recovered short-timescale slopes (squares in Figure~\ref{fig:sf_ngc4395}) are mostly steeper than the DRW expectation, with $\beta$ ranging from 1.44 to 2.16 (median 1.75). The TESS dataset is an exception, remaining consistent with the DRW, but its shortest accessible timescale is $\simeq 10^3$~s, an order of magnitude longer than the $\simeq 10^2$~s reached by the others. The steepening appears across different nights and telescopes, ruling out a single-dataset origin. However, due to the degeneracy discussed in Section~\ref{sec:k16_degeneracy}, the slopes are poorly constrained, and the evidence for a departure from DRW at short timescales in NGC~4395 remains suggestive rather than definitive.

A finite realization of a stochastic process can inherently deviate from its long-term expectation \citep{Kozlowski2021AcA....71..103K}. Establishing the statistical significance of the observed steepening will require simulations matched to the cadence and baseline of each observation, once a reliable intrinsic SF is recovered. Future surveys such as the Wide Field Survey Telescope \citep[WFST;][]{WFST2023SCPMA..6609512W} and the Legacy Survey of Space and Time \citep[LSST;][]{Ivezic2019} will provide larger samples of NGC~4395-like IMBH candidates (e.g., \citealt{Wang2023MNRAS}, \citealt{Guo2026arXiv260902179G}) with improved photometric calibration, enabling the intrinsic variability properties of IMBHs to be characterized as a population.

\section{Conclusion} \label{sec:conclusion}

Whether weak intrinsic variability in AGNs can be reliably recovered depends on how the contribution from the photometric noise is removed. Comparing three SF estimators across a wide range of signal-to-noise ratio, we find that all recover the intrinsic SF where the signal exceeds the noise, but fail differently once it does not: pairwise subtraction (P92), when negative variance estimates are discarded as is common in SF-amplitude measurements, produces an artificial floor (Section~\ref{sec:p92_floor}), the parametric fit (K16) becomes degenerate between noise and weak signal (Section~\ref{sec:k16_degeneracy}), and only the Bayesian estimator (S17) remains nearly unbiased above the detection limit and returns honest upper limits below it (Section~\ref{sec:s17_optimal}). 
These conclusions remain valid for non-DRW processes (Section~\ref{sec:beyond_drw}). Both P92 and S17 further require photometric errors accurate enough that they must be calibrated against non-variable stars (Section~\ref{sec:photo_errors}).

Reliable measurement of weak AGN variability therefore requires well-calibrated photometric errors and sufficient informative pairs, together with an estimator that accounts for the noise and the uncertainty in the intrinsic signal.
These considerations will matter increasingly for upcoming time-domain surveys such as WFST \citep{WFST2023SCPMA..6609512W} and LSST \citep{Ivezic2019}, which will deliver large samples of AGNs monitored with unprecedented cadence and photometric precision. They will set the reliability of intrinsic variability measurements from low-luminosity nuclei to extremely stable quasars, and for newly identified populations such as the JWST little red dots.

\begin{acknowledgments}
We are grateful to the anonymous referee and statistics editor for constructive comments.
This work is supported by the National Key R\&D Program of China (grant No. 2023YFA1608100), the National Natural Science Foundation of China (grant Nos. 12373016 and 12533006), and the Cyrus Chun Ying Tang Foundations, which also provided a special fellowship to Z.B.S.
L.F. and J.X.W. gratefully acknowledge support from the Guizhou Provincial Major Scientific and Technological Programs XKBF (2025)010 and XKBF (2025)011. 

\end{acknowledgments}

\bibliography{sample701}{}
\bibliographystyle{aasjournalv7}


\end{CJK*}
\end{document}